\documentclass[12pt]{article}

\usepackage{graphics,cite,amssymb,epsfig,float,psfrag,axodraw}
\usepackage[usenames,dvips]{color}
\usepackage{amsmath}
\usepackage{rotating}
\usepackage{multicol}
\usepackage{bigcenter}

\newcommand{\beq}{\begin{eqnarray}} 
\newcommand{\eeq}{\end{eqnarray}}

\newcommand{\lum}{\mbox{$\int\! \cal L$}}

\newcommand{\gghh}{\ensuremath{gg \rightarrow HH}}

\newcommand{\mh}{\ensuremath{M_{H}}}
\newcommand{\mhh}{\ensuremath{M_{HH}}}
\newcommand{\pth}{\ensuremath{P_{T,H}}}
\newcommand{\thetas}{\ensuremath{\theta^{\star}_{HH}}}

\newcommand{\etah}{\ensuremath{\eta_{H}}}

\newcommand{\bbtt}{\ensuremath{b\bar{b} \tau\bar{\tau}}}
\newcommand{\bbww}{\ensuremath{b\bar{b}  W^+W^-}}
\newcommand{\bbgg}{\ensuremath{b\bar{b} \gamma\gamma}}

\newcommand{\Lep}{\ensuremath{\ell}}

\begin{document}

\vspace{.8cm}

\hfill  KA--TP--44--2012

\hfill SFB/CPP--12--102

\hfill  LPT--ORSAY--12--124

\hfill PSI--PR--12--10

\vspace*{1.4cm}

\begin{center}

{\large\bf The measurement of the Higgs self--coupling at the LHC:}
 
\vspace*{2mm} 

{\large\bf theoretical status}

\vspace*{.6cm}

{\sc J. Baglio$^1$, A. Djouadi$^{2}$, R. Gr\"ober$^1$,}

\vspace*{2mm}

{\sc M.M. M\"uhlleitner$^1$, J. Quevillon$^2$} and {\sc M. Spira$^3$}

\vspace*{.8cm}

\begin{small}

$^1$ Institut f\"ur Theoretische Physik, KIT, D-76128 Karlsruhe, Germany.

\vspace*{1mm}

$^2$ Laboratoire de Physique Th\'eorique, U. Paris--Sud and CNRS, F--91405
Orsay, France.

\vspace*{1mm}

$^3$Paul Scherrer Institute, CH-5232 Villigen PSI, Switzerland.

\end{small}

\end{center}

\vspace*{1cm}

\begin{abstract} 

Now that the Higgs boson has been observed by the ATLAS and CMS
experiments at the LHC, the next important step would be to measure
accurately its properties to establish the details of the electroweak
symmetry breaking mechanism. Among the measurements which need to be
performed, the determination of the Higgs self--coupling in processes
where the Higgs boson is produced in pairs is of utmost importance. In
this paper, we discuss the various processes which allow for the
measurement of the trilinear Higgs coupling: double Higgs production
in gluon fusion, vector boson fusion, double Higgs--strahlung and
associated production with a top quark pair. We first evaluate the
production cross sections for these processes at the LHC with
center--of--mass energies ranging from the present $\sqrt s=8$~TeV to
$\sqrt s=100$ TeV, and discuss their sensitivity to the trilinear
Higgs coupling. We include the various higher order QCD radiative
corrections, at next--to--leading order for gluon and vector boson
fusion and at next--to--next--to--leading order for associated double
Higgs production with a gauge boson. The theoretical uncertainties on
these cross sections are estimated. Finally, we discuss the various
channels which could allow for the detection of the double Higgs
production signal at the LHC and estimate their potential to probe the
trilinear Higgs coupling.

\end{abstract} 

\newpage

\section{Introduction} 

A bosonic particle with a mass of about 125~GeV has been observed by
the ATLAS and CMS Collaborations at the LHC~\cite{LHC} and it has,
{\it grosso modo}, the properties of the long sought Higgs particle
predicted in the Standard Model (SM)~\cite{Higgs}. This closes the
first chapter of the probing of the mechanism that triggers the
breaking of the electroweak symmetry and generates the fundamental
particle masses. Another, equally important chapter is now opening:
the precise determination of the properties of the produced
particle. This is of extreme importance in order to establish that
this particle is indeed the relic of the mechanism responsible for the
electroweak symmetry breaking and, eventually, to pin down effects of
new physics if additional ingredients beyond those of the SM are
involved in the symmetry breaking mechanism. To do so, besides
measuring the mass, the total decay width and the spin--parity quantum
numbers of the particle, a precise determination of its couplings to
fermions and gauge bosons is needed in order to verify the fundamental
prediction that they are indeed proportional to the particle
masses. Furthermore, it is necessary to measure the Higgs
self--interactions. This is the only way to reconstruct the scalar
potential of the Higgs doublet field $\Phi$, that is responsible for
spontaneous electroweak symmetry breaking,
\beq
V_H= \mu^2 \Phi^\dagger\Phi + \frac12 \lambda  (\Phi^\dagger\Phi)^2 \
; \ \ \  \lambda = \frac{M_H^2}{v^2}\ {\rm and}  \ \mu^2= - \frac12
M_H^2 \; ,
\eeq
with $v= 246$ GeV. Rewriting the Higgs potential in terms of a
physical Higgs boson leads to the trilinear Higgs self--coupling
$\lambda_{HHH}$, which in the SM is uniquely related to the mass of
the Higgs boson,
\beq
\lambda_{HHH}= \frac{3 M_H^2}{v}\, . 
\eeq
This coupling is only accessible in double Higgs
production~\cite{tripleh1a,tripleh1b,tripleh2,tripleh3}. One thus
needs to consider the usual channels in which the Higgs boson is
produced singly~\cite{Review}, but allows for the state to be off
mass--shell and to split up into two real Higgs bosons. At hadron
colliders, four main classes of processes have been advocated for
Higgs pair production:
\vspace*{-2mm}
\begin{itemize}
\item[$a)$] the gluon fusion mechanism, $gg \to HH$, which is
mediated by loops of heavy quarks (mainly top quarks)
that couple strongly to the Higgs
boson~\cite{pp-ggHH-LO00,pp-ggHH-LO0,pp-ggHH-LO,pp-ggHH-LO1};

\vspace*{-2mm}
\item[$b)$] the $WW/ZZ$ fusion processes (VBF), $qq' \to V^* V^* qq'
 \to HHqq'$ ($V=W, Z$), which lead to two Higgs particles and two jets
 in the final state~\cite{pp-ggHH-LO00,pp-VVHH,pp-VVH-Abas};

\vspace*{-2mm}
\item[$c)$] the double Higgs--strahlung process, $q\bar{q}' \to V^*
 \to VHH$ ($V=W,Z$), in which the Higgs bosons are radiated from either a
 $W$ or a $Z$ boson~\cite{pp-HHV};

\vspace*{-2mm}
\item[$d)$] associated production of two Higgs bosons with a top quark
 pair, $pp \to t\bar t HH$~\cite{pp-HHtt}.
\end{itemize}
As they are of higher order in the electroweak coupling and the phase
space is small due to the production of two heavy particles in the
final state, these processes have much lower production cross
sections, at least two orders of magnitude smaller, compared to the
single Higgs production case. In addition, besides the diagrams with
$H^* \to HH$ splitting, there are other topologies which do not
involve the trilinear Higgs coupling, e.g. with both Higgs bosons
radiated from the gauge boson or fermion lines, and which lead to the
same final state. These topologies will thus dilute the dependence of
the production cross sections for double Higgs production on the
$\lambda_{HHH}$ coupling. The measurement of the trilinear Higgs
coupling is therefore an extremely challenging task and very high
collider luminosities as well as high energies are required. We should
note that to probe the quadrilinear Higgs coupling, $\lambda_{HHHH} =
3M_H^2/v^2$, which is further suppressed by a power of $v$ compared
to the triple Higgs coupling, one needs to consider triple Higgs
production processes~\cite{tripleh1a,gg-HHH}. As their cross sections
are too small to be measurable, these processes are not viable in a
foreseen future so that the determination of this last coupling seems
hopeless.

In Ref.~\cite{tripleh2}, the cross sections for the double Higgs
production processes and the prospects of extracting the Higgs
self-coupling have been discussed for the LHC with a 14 TeV
center--of--mass (c.m.) energy in both the SM and its minimal
supersymmetric extension (MSSM) where additional channels occur in the
various processes.

In the present paper, we update the previous analysis. In a sense, the
task is made easier now that the Higgs boson mass is known and can be
fixed to $M_H\approx 125$ GeV. However, lower c.m. energies have to be
considered such as the current one, $\sqrt{s}=8$~TeV. In addition,
there are plans to upgrade the LHC which could allow to reach
c.m.~energies of about 30~TeV~\cite{33tev-cern} and even up to
100~TeV. These very high energies will be of crucial help to probe
these processes.

Another major update compared to Ref.~\cite{tripleh2} is that we will
consider all main processes beyond leading order (LO)
in perturbation theory, i.e. we will implement the important higher
order QCD corrections. In the case of the gluon fusion mechanism,
$gg\to HH$, the QCD corrections have been calculated at
next-to--leading-order (NLO) in the low~energy limit in
Ref.~\cite{pp-ggHH-NLO}. They turn out to be quite large, almost
doubling the production cross section at $\sqrt s=14$ TeV, in much the
same manner as for single Higgs production~\cite{ggH-NLO}. In fact,
the QCD corrections for single and double Higgs productions are
intimately related and one should expect, as in the case of $gg\to H$,
a further increase of the total cross section by $\approx 30\%$ once
the next-to-next-to-leading (NNLO) corrections are also
included~\cite{ggH-NNLO0, ggH-NNLO}.

It is well known that for single Higgs production in the vector boson
fusion process $qq'\to Hqq'$ there is no gluon exchange between the
two incoming/outgoing quarks as the initial and final quarks are in
color singlet states at LO. Then the NLO QCD corrections consist
simply of the known corrections to the structure
functions~\cite{VVH-NLO,VHH-VBFNLO, VVH-NLO2}. The same can be said in
the case of double Higgs production $qq'\to HHqq'$, and in this paper
we will implement the NLO QCD corrections to this process in the
structure function approach. The NNLO corrections in this approach
turn out to be negligibly small for single Higgs
production~\cite{VVH-NNLO} and we will thus ignore them for double
Higgs production.

In the single Higgs--strahlung process, $q\bar{q}' \to V^*\to VH$, the
NLO QCD corrections can be inferred from those of the Drell--Yan
process $q\bar{q}' \to V^*$~\cite{HV-NLO}. This can be extended to
NNLO~\cite{ggH-NNLO0,HV-DY,HV-NNLO} but, in the case of $ZH$
production, one needs to include the $gg$ initiated contribution,
$gg\to ZH$~\cite{HV-gg,HV-NNLO} as well as some additional subleading 
corrections~\cite{HV-sublead}. The same is also true for double
Higgs--strahlung and we will include in this paper the Drell--Yan part
of the corrections up to NNLO. In the case of $ZHH$ final states, we
will determine the additional contribution of the pentagon
diagram $gg \to ZHH$ which turns out to be quite substantial,
increasing the total cross section by up to $30\%$ at $\sqrt s =14$
TeV.

In the case of the $pp\to t \bar{t} HH$ process, the determination of
the cross section at LO is already rather complicated. We will not
consider any correction beyond this order (that, in any case, has not
been calculated) and just display the total cross section  without
further analysis. We simply note that the QCD corrections in the
single Higgs case, $pp \to t\bar{t} H$, turn out to be quite
modest. At NLO, they are small at $\sqrt s=8$ TeV and increase the
cross section by less than $\approx 20\%$ at $\sqrt s=14$
TeV~\cite{Htt-NLO}. We also note that this channel is plagued by huge
QCD backgrounds.

In addition, the electroweak radiative corrections to these double
Higgs production processes have not been calculated yet. Nevertheless,
one expects that they are similar in size to those affecting the
single Higgs production case, which are at the few percent level at
the presently planned LHC c.m. energies~\cite{Gambino,VBF-EW, VH-EW}
(see also Ref.~\cite{LHCXS} for a review). They should thus not affect
the cross sections in a significant way and we will ignore this issue
in our analysis.

After determining the $K$--factors, i.e. the ratios of the higher order
to the lowest order cross sections consistently evaluated with the
value of the strong coupling $\alpha_s$ and the parton distribution
functions taken at the considered perturbative order, a next step will
be to estimate the theoretical uncertainties on the production cross
sections in the various processes. These stem from the variation of
the renormalization and factorization scales that enter the processes
(and which gives a rough measure of the missing higher order
contributions), the uncertainties in the parton distribution functions
(PDF) and the associated one on the strong coupling constant
$\alpha_s$ and, in the case of the $gg\to HH$ process, the uncertainty
from the use of an effective approach with an infinitely heavy virtual
top quark, to derive the NLO corrections (see also
Ref.~\cite{Gillioz:2012se}). This will be done in much the same way as
for the more widely studied single Higgs production
case~\cite{LHCXS,JB}.

Finally, we perform a preliminary analysis of the various channels in
which the Higgs pair can be observed at the LHC with a c.m.~energy of
$\sqrt s=14$ TeV assuming up to 3000~fb$^{-1}$ collected data, and
explore their potential to probe the $\lambda_{HHH}$
coupling. Restricting ourselves to the dominant $gg \to HH$ mechanism
in a parton level approach\footnote{Early and more recent parton level
  analyses of various detection channels have been performed in
  Refs.~\cite{early,bbtautau,bbWW} with the recent ones heavily
  relying on jet--substructure techniques~\cite{boost}. However, a
  full and realistic assessment of the LHC to probe the trilinear
  coupling would require the knowledge of the exact experimental
  conditions with very high luminosities and a full simulation of the
  detectors which is beyond the scope of this paper. Only the ATLAS
  and CMS Collaborations are in a position to perform such detailed
  investigations and preliminary studies have already
  started~\cite{atlas-cracow}.}, we first discuss the kinematics of the
process, in particular the transverse momentum distribution of the
Higgs bosons and their rapidity distribution at leading order. We then
evaluate the possible cross sections for both the signal and the major
backgrounds. As the Higgs boson of a mass around $125$ GeV dominantly
decays into $b$--quark pairs with a branching ratio of $\approx 60\%$
and other decay modes such as $H\to \gamma\gamma$ and $H\to WW^* \to
2\ell 2\nu$ are rare~\cite{hdecay}, and as the production cross
sections are already low, we will focus on the three possibly
promising detection channels $gg\to HH \to b\bar b \gamma \gamma,
b\bar b \tau \bar{\tau}$ and $b\bar b W^+ W^-$. Very high luminosities,
${\cal O}$(ab$^{-1}$) would be required to have some sensitivity
on the $\lambda_{HHH}$ coupling.

The rest of the paper is organized as follows. In the next section, we
discuss the QCD radiative corrections to double Higgs production in
the gluon fusion, vector boson fusion and Higgs--strahlung processes
(the $t\bar{t}HH$ process will be only considered at tree--level) and
how they are implemented in the programs {\tt HPAIR}~\cite{hpair}, 
{\tt VBFNLO}~\cite{vbfnlo} and a code developed by us to evaluate the
inclusive cross sections in Higgs--strahlung processes. In section 3,
we evaluate the various theoretical uncertainties affecting these
cross sections and collect at $M_H=125$ GeV the double Higgs
production cross sections at the various LHC energies. We also study
the sensitivity in the different processes to the trilinear Higgs
self--coupling. Section~4 will be devoted to a general discussion of
the channels that could allow for the detection of the two Higgs boson
signal at a high--luminosity 14 TeV LHC, concentrating on the dominant
$gg \to HH$ process, together with an analysis of the major
backgrounds. A short conclusion is given in the last section.

\section{Higgs pairs at higher orders in QCD}

Generic Feynman diagrams for the four main classes of processes
leading to double Higgs production at hadron colliders, gluon fusion,
$WW/ZZ$ fusion, double Higgs--strahlung and associated production with
a top quark pair, are shown in Fig.~1. As can be seen in each process,
one of the Feynman diagrams involves the trilinear Higgs boson coupling,
$\lambda_{HHH}=3 M_H^2/v$, which can thus be probed in principle. The
other diagrams involve the couplings of the Higgs boson to fermions
and gauge bosons and are probed in the processes where the Higgs
particle is produced singly.

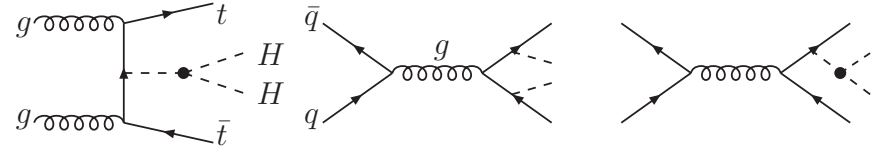
\begin{figure}[h!]
\SetScale{0.75}
\vspace*{-4mm}
\begin{picture}(100,80)(-30,10)
\hspace*{5mm}
\SetWidth{0.9}
\put(-45, 70){\underline{(a) $gg$ double-Higgs fusion: $gg\to HH$}}
\Vertex(120,30){3}
\Gluon(0,60)(40,60){3.5}{5.5}
\Gluon(0,0)(40,0){3.5}{5.5}
\ArrowLine(40,60)(80,30)
\ArrowLine(80,30)(40,0)
\ArrowLine(40,0)(40,60)
\DashLine(80,30)(120,30){5}
\DashLine(120,30)(150,60){5}
\DashLine(120,30)(150,0){5}
\put(116,45){$H$}
\put(116,-4){$H$}
\put(72,26){$H$}
\put(-11,-2){$g$}
\put(-11,45){$g$}
\put(37,22){$Q$}
\hspace*{7cm}
\Gluon(0,0)(40,0){3.5}{5.5}
\Gluon(0,60)(40,60){3.5}{5.5}
\ArrowLine(40,60)(90,60)
\ArrowLine(90,0)(40,0)
\ArrowLine(40,0)(40,60)
\ArrowLine(90,60)(90,0)
\DashLine(90,60)(130,60){5}
\DashLine(90,0)(130,0){5}
\put(101,45){$H$}
\put(101,-2){$H$}
\put(-11,-2){$g$}
\put(-11,45){$g$}
\put(37,22.5){$Q$}
\end{picture}

\vspace*{0mm}
\begin{center}
\hspace*{-14cm}
\SetWidth{0.9}
\begin{picture}(300,80)(0,25)
\hspace*{1.8cm}
\ArrowLine(150,25)(195,25)
\ArrowLine(150,75)(195,75)
\ArrowLine(195,25)(240,15)
\ArrowLine(195,75)(240,85)
\Photon(195,25)(195,75){3.5}{5.5}
\DashLine(195,50)(225,50){4}
\DashLine(225,50)(255,60){4}
\DashLine(225,50)(255,40){4}
\put(70, 80){\underline{(b) $WW/ZZ$ double-Higgs fusion: $qq'\to
    HHqq'$}}
 \Vertex(225, 50){3}
\Text(106,22.5)[]{$q$}
\Text(106,52.5)[]{$q'$}
\Text(185,11)[]{$q$}
\Text(187,65)[]{$q'$}
\Text(158,49)[]{$V^*$}
\Text(158,26)[]{$V^*$}
\Text(203,45)[]{$H$}
\Text(203,30)[]{$H$}
\ArrowLine(295,25)(340,25)
\ArrowLine(295,75)(340,75)
\ArrowLine(340,25)(385,15)
\ArrowLine(340,75)(385,85)
\DashLine(340,60)(385,65){4}
\Photon(340,25)(340,75){3.5}{6.5}
\DashLine(340,40)(385,35){4}
%
\ArrowLine(425,25)(470,25)
\ArrowLine(425,75)(470,75)
\ArrowLine(470,25)(515,15)
\ArrowLine(470,75)(515,85)
\DashLine(470,50)(515,65){4}
\Photon(470,25)(470,75){3.5}{6.5}
\DashLine(470,50)(515,35){4}
\end{picture}
\vspace*{-3.mm}
\end{center}
\begin{center}
\hspace*{-14cm}
\vspace*{0cm}
\SetWidth{0.9}
\begin{picture}(300,100)(-50,10)
\ArrowLine(150,25)(185,50)
\ArrowLine(185,50)(150,75)
\Photon(185,50)(230,50){3.5}{5.5}
\Photon(230,50)(265,25){3.5}{5.5}
\DashLine(230,50)(250,60){4}
\DashLine(250,60)(265,75){4}
\DashLine(250,60)(265,45){4}
 \Vertex(250, 60){3}
\put(70, 75){\underline{(c) Double Higgs-strahlung: $q\bar{q}'\to
    ZHH/WHH$}}
\Text(107,22.5)[]{$q$}
\Text(107,53)[]{$\bar{q}'$}
\Text(158,49)[]{$V^*$}
\Text(206,22.5)[]{$V$}
\Text(206,56)[]{$H$}
\Text(206,37.5)[]{$H$}
\ArrowLine(295,25)(330,50)
\ArrowLine(330,50)(295,75)
\DashLine(375,50)(410,25){4}
\Photon(375,50)(410,75){3.5}{5.5}
\Photon(330,50)(375,50){3.5}{5.5}
\DashLine(390,55)(410,45){4}
\vspace*{3mm}
\ArrowLine(445,25)(480,50)
\ArrowLine(480,50)(445,75)
\Photon(480,50)(525,50){3.5}{5.5}
\DashLine(525,50)(560,40){4}
\Photon(525,50)(560,75){3.5}{4.5}
\DashLine(525,50)(560,25){4}
\end{picture}
\vspace*{9.mm}
\end{center}
\begin{center}
\hspace*{-14cm}
\SetWidth{0.9}
\begin{picture}(300,100)(-20,-30)
\hspace*{1cm}
\Gluon(150,25)(195,25){3.5}{5.5}
\Gluon(150,75)(195,75){3.5}{5.5}
\ArrowLine(240,15)(195,25)
\ArrowLine(195,75)(240,85)
\ArrowLine(195,25)(195,75)
\DashLine(195,50)(225,50){4}
\DashLine(225,50)(255,60){4}
\DashLine(225,50)(255,40){4}
\Vertex(225,50){3}
\Text(109,20)[]{$g$}
\Text(109,55)[]{$g$}
\Text(184,15)[]{$\bar t$}
\Text(184,60)[]{$t$}
\Text(202.5,45)[]{$H$}
\Text(202.5,30)[]{$H$}
\ArrowLine(295,25)(330,50)
\ArrowLine(330,50)(295,75)
\DashLine(390,60.71)(410,55){4}
\ArrowLine(375,50)(410,75)
\ArrowLine(410,25)(375,50)
\Gluon(330,50)(375,50){3.5}{5.5}
\DashLine(390,39.25)(410,45){4}
\put(215, 19){$q$}
\put(215, 56){$\bar{q}$}
\put(264, 45){$g$}
\vspace*{3mm}
\ArrowLine(445,25)(480,50)
\ArrowLine(480,50)(445,75)
\Gluon(480,50)(525,50){3.5}{5.5}
\DashLine(540,61.75)(570,39.2811){4}
\DashLine(555,50)(570,60){4}
\Vertex(555,50){3}
\ArrowLine(525,50)(560,75)
\ArrowLine(560,25)(525,50)
\put(70, 85){\underline{(d) Associated production with top-quarks:
    $q\bar{q}/gg\to t\bar{t}HH$}}
 \end{picture}\vspace*{3.mm}
\end{center}
\vspace*{-2.6cm}
\hspace*{-4cm}
\it{\caption{Some generic Feynman diagrams contributing to Higgs pair
    production at hadron colliders.\label{processeslo}}}
\vspace*{-2mm}
\end{figure}

In this section we will discuss the production cross sections for the
first three classes of processes, including the higher order QCD
corrections. We will first review the gluon channel and then we will
move on to the higher--order corrections in the weak boson fusion and
Higgs--strahlung channels.

\subsection{The gluon fusion process}

The gluon fusion process is -- in analogy to single Higgs production
-- the dominant Higgs pair production process. The cross section is
about one order of magnitude larger than the second largest process
which is vector boson fusion. As can be inferred from
Fig.~\ref{processeslo}a) it is mediated by loops of heavy quarks
which in the SM are mainly top quarks. Bottom quark loops contribute
to the total cross section with less than $1\%$ at LO.
\par
The process is known at NLO QCD in an effective field theory (EFT)
approximation by applying the low energy theorem
(LET)~\cite{ggH-NLO,Ellis:1975ap,Kniehl:1995tn} which means that
effective couplings of the gluons to the Higgs bosons are obtained by
using the infinite quark mass approximation. The hadronic cross
section at LO is given by
\begin{equation}
  \sigma_{\mathrm{LO}}  = \int_{\tau_0}^1 d\tau~
  \hat\sigma_{\mathrm{LO}}(\hat{s} = \tau s)~ \int_{\tau}^1
  \frac{dx}{x}  f_g(x; \mu_F^2)f_{g}\left(\frac{\tau}{x};
    \mu_F^2\right)\, ,
\end{equation}
with $s$ being the hadronic c.m. energy, $\displaystyle
\tau_0=4 M_H^2/s$, and $f_g$ the gluon distribution function taken at
a typical scale $\mu_F$ specified below. The partonic cross section at
LO, $\hat{\sigma}_{\mathrm{LO}}$, can be cast into the form
\begin{equation}
  \hat \sigma_{\mathrm{LO}}(gg\to HH) = \int_{\hat t_-}^{\hat t_+}
  d\hat t \, \frac{G_F^2 \alpha_s^2(\mu_R)}{256 (2\pi)^3} \left\{
    \left| \frac{\lambda_{HHH}\;v}{\hat{s}-M_H^2+iM_H\Gamma_H}
      F_\triangle + F_\Box \right|^2 + \left|  G_\Box \right|^2
  \right\}\, ,
\end{equation}
where
\begin{equation}
\hat{t}_{\pm}=-\frac{\hat{s}}{2}\left(1-2\frac{M_{H}^2}{\hat{s}}\mp\sqrt{1-\frac{4
      M_{H}^2}{\hat{s}}}\right)\, ,
\end{equation}
with $\hat{s}$ and $\hat{t}$ denoting the partonic Mandelstam
variables. The triangular and box form factors $F_\triangle$, $F_\Box$
and $G_\Box$ approach constant values in the infinite top quark mass
limit,
\begin{equation}
  F_\triangle\to \frac{2}{3},\hspace*{0.5cm}F_\Box\to -\frac{2}{3},
  \hspace*{0.5cm}G_\Box\to 0\;.
\end{equation}
The expressions with the complete mass dependence are rather lengthy
and can be found in Ref.~\cite{pp-ggHH-LO1} as well as the NLO QCD
corrections in the LET approximation in Ref.~\cite{pp-ggHH-NLO}.
\par
The full LO expressions for $F_\triangle, F_\Box$ and $G_\Box$ are
used wherever they appear in the NLO corrections in order to improve
 the perturbative results, similar to what
has been done in the single Higgs production case where using the
exact LO expression reduces the disagreement between the full NLO
result and the LET result~\cite{Review,ggH-NLO}.
\par

For the numerical evaluation we have used the publicly available
code {\tt HPAIR}~\cite{hpair} in which the known NLO corrections are
implemented. As a central scale for this process we choose
\begin{equation}
  \mu_0=\mu_{R}=\mu_{F}= M_{HH}\;,
\label{ggHH-scalechoice}
\end{equation}
where $M_{HH}$ denotes the invariant mass of the Higgs boson
pair. This is motivated by the fact that the natural scale choice in
the process $gg\to H$ is $\mu_0=M_H$. Extending this to Higgs pair
production naturally leads to the scale choice of
Eq.~(\ref{ggHH-scalechoice}). The motivation to switch to $\mu_0=1/2\,
M_H$ in single Higgs production comes from the fact that it is a way
to acccount for the $\sim +10\%$ next--to--next--to--leading
logarithmic (NNLL) corrections~\cite{LHCXS, ggH-NNLL} in a fixed order
NNLO calculation. It also improves the perturbative convergence from
NLO to NNLO~\cite{ggH-scale}. Still NNLO and NNLL calculations for
$gg\to HH$ process are not available at the moment, not to mention an
exact NLO calculation that would be the starting point of further
improvements. It then means that there is no way to check wether these
nice features appearing in single Higgs production when using
$\mu_0=1/2\, M_H$ would still hold in the case of Higgs pair
production when using $\mu_0=1/2\, M_{HH}$. We then stick to the scale
choice of Eq.~(\ref{ggHH-scalechoice}). The $K$--factor, describing
the ratio of the cross section at NLO using NLO PDFs and NLO
$\alpha_s$ to the leading order cross section consistently evaluated
with LO PDFs and LO $\alpha_s$, for this process is
\begin{equation}
  K\sim2.0\; (1.5) \hspace*{0.5cm} \text{for } \sqrt{s}=8\;(100)\;
  \text{TeV}\;.
  \label{ggHH_kfactor}
\end{equation}

\subsection{The vector boson fusion process}

The structure of the Higgs pair production process through vector boson
fusion~\cite{pp-VVHH} is very similar to the single Higgs production
case. The vector boson fusion process can be viewed as the double
elastic scattering of two (anti)quarks with two Higgs bosons radiated
off the weak bosons that fuse. In particular this means that the
interference with the double Higgs--strahlung process $qq'\to V^* HH\to
qq' HH$ is negligible and this latter process is treated separately. This is
justified by the kinematics of the process with two widely separated
quark jets of high invariant mass and by the color flow of the
process. This leads to the structure function approach that has been
applied with success to calculate the QCD corrections in the vector
boson fusion production of a single Higgs boson~\cite{VVH-NLO,
  VHH-VBFNLO, VVH-NNLO, VVH-NLO2}.
\begin{figure}[h!]
\begin{center}
\includegraphics[scale=0.52]{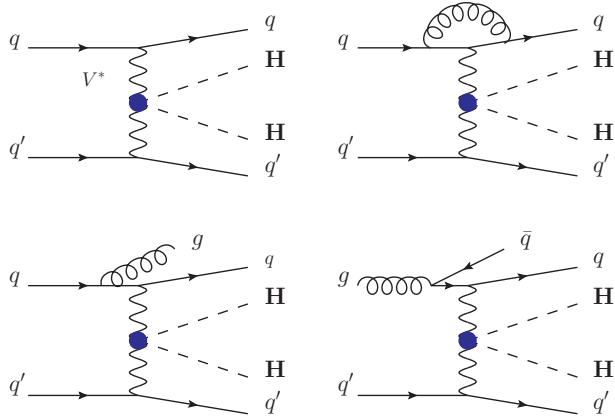}
\end{center}
\it{\vspace{-6mm}\caption{Generic diagrams contributing to the NLO
    corrections to $qq'\to HH qq'$. Shown are the LO diagram (upper
    left) and the NLO corrections for the upper quark line. The blob
    of the $VVHH$ vertex is a shortcut for the three diagrams shown in
    Figs.~\ref{processeslo}b) and~\ref{VVHH-tensor}.\label{VVHH-NLO}}}
\end{figure}
Generic diagrams contributing at NLO QCD order are shown in
Fig.~\ref{VVHH-NLO}. For simplicity only the diagrams with the QCD
corrections to the upper quark line are shown. The calculation
involving the second quark line is identical. The blob of the vertex
$VVHH$ is a shortcut for the diagrams depicted in
Fig.~\ref{VVHH-tensor}, which include charged currents (CC) with
$W^{\pm}$ bosons and neutral currents (NC) with a $Z$ boson
exchange. As can be seen only one of the three diagrams involves the
trilinear Higgs coupling. The other diagrams act as irreducible
background and lower the sensitivity of the production process to
the Higgs self--coupling.
\begin{figure}[h!]
\begin{center}
\includegraphics[scale=0.5]{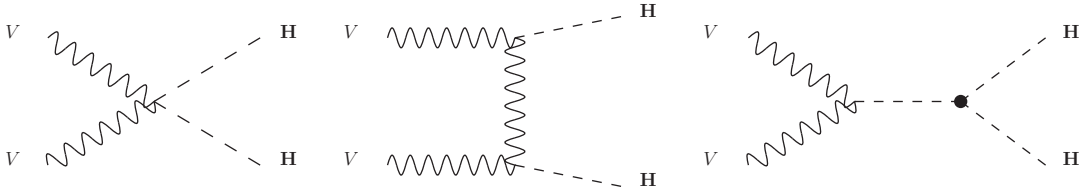}
\end{center}
\it{\vspace{-6mm}\caption{Diagrams contributing to the $VVHH$
    vertex.\label{VVHH-tensor}}}
\end{figure}

We have calculated the NLO QCD corrections in complete analogy to
the single Higgs VBF process~\cite{VHH-VBFNLO}. The real emission 
contributions are given by a gluon attached to the quark lines
either in the initial or the final state and from the gluon--quark
initial state. As we are working in the structure function approach,
the corrections of the upper and lower quark lines do not interfere
and are simply added incoherently. The amplitudes have the following
structure,
\begin{equation}
\mathcal{A}_{H\! Hqq'} \propto T^{\mu\nu}_{V^*V^*} J^{q}_{\mu}
J^{q'}_\nu\, ,
\end{equation}
where $T^{\mu\nu}_{V^*V^*}$ stands for the tensor structure of the
diagrams depicted in Fig.~\ref{VVHH-tensor} and $J^{q,q'}_\mu$ are the
quark currents of the upper and lower lines, respectively, with
four-momenta $q$, $q'$. The calculation is done
numerically using the {\it Catani-Seymour} dipole subtraction
method~\cite{dipole} to regularize the infrared divergencies. The
formulae for the subtraction terms as well as the finite corrections
are identical to the ones for single Higgs VBF production as only the
quark currents are involved. They can be found in
Ref.~\cite{VHH-VBFNLO}.

We have implemented this calculation in the {\tt VBFNLO}
code~\cite{vbfnlo} in which we have provided the tensor structure
depicted in Fig.~\ref{VVHH-tensor} which has been calculated with
MadGraph~\cite{madgraph}. Up to now the {\tt VBFNLO} implementation
only involves on--shell Higgs pairs. We have found an increase of
$\sim +7\%$ of the total cross section compared to the LO result when
using the central scale
\begin{equation}
\mu_0 = \mu_R=\mu_F=Q_{V^*},
\end{equation}
with $Q_{V^*}$ being the momentum of the exchanged weak bosons ($V^* =
W^*,Z^*$)\footnote{In order to stay within the perturbative regime a
  cut $Q_{V^*} \geq 2$ GeV has to be imposed, see
  Ref.~\cite{VVH-NLO}.}. This result is in agreement with a previous
calculation done in the context of the two Higgs doublet
model~\cite{THDM-Figy}.

\subsection{The Higgs--strahlung process}

The production of a Higgs pair in association with a vector boson has
been calculated for the first time quite a while ago~\cite{pp-HHV} and
shares common aspects with the single Higgs--strahlung process. The
NLO corrections can be implemented in complete analogy to single
Higgs-strahlung~\cite{HV-NLO}. We will update in this
paper the former results and present the NNLO corrections to the $WHH$
and $ZHH$ inclusive production cross sections. These calculations have
been implemented in a code which shall become publicly available.

At LO the process $p p \to V HH$ ($V=W,Z$) is given by
quark--antiquark annihilations in $s$--channel mediated processes
involving three Feynman diagrams, see Fig.\;\ref{processeslo}c). As
can be seen only one of the three diagrams involves the trilinear
Higgs coupling. The sensitivity to this coupling is then diluted by
the remaining diagrams. After integrating over the azimuthal angle we
are left with the following partonic differential cross section with
$\hat{s}$ being the partonic c.m. energy (see also
Ref.~\cite{tripleh2}),
\begin{eqnarray}
 \hspace{-8mm} \frac{d \hat{\sigma}_{VHH}^{\rm LO}}{dx_1 dx_2} & \!\!=\!\! &
 \displaystyle \frac{G_F^3 M_V^6(a_q^2+v_q^2)}{1149\sqrt{2} \pi^3 \hat{s}
   (1-\mu_V^2)} \left[ \frac{1}{8} f_0 C_{HHH}^2 +
   \frac{1}{4\mu_V(1-x_1+\mu_H-\mu_V)} \times \right. \nonumber \\
  & & \nonumber \\
  & & \displaystyle \left( \frac{f_1}{1-x_1+\mu_H-\mu_V} \left. +
      \frac{f_2}{1-x_2+\mu_H-\mu_V}+2\mu_V f_3 C_{HHH}\right) +
    \{x_1\leftrightarrow x_2\} \right],\, 
  \label{vhh-partonic} 
\end{eqnarray}
where we use of the following notation,
\begin{equation}
  \mu_V = \frac{M_V^2}{\hat{s}},\,\,\,\, \mu_H =
  \frac{M_H^2}{\hat{s}},\,\,\,\, 
  x_1 = \frac{2 E_H}{\sqrt{\hat{s}}},\,\,\,\, x_2 = \frac{2
    E_V}{\sqrt{\hat{s}}},
\end{equation}
and the reduced couplings of the quarks to the vector bosons,
$a_q=v_q=\sqrt{2}$ for $V=W$ and any quark $q$, $a_u=1$ and $v_u=1-8/3
\sin^2 \theta_W$ for $q=u, s$ and $V=Z$, $a_d=-1$ and
$v_d=-1+4/3\sin^2\theta_W$ for $q=d,c,b$ and $V=Z$. The coefficients
$f_i$ as well as $C_{HHH}$ are
\begin{eqnarray}
  f_0 & = & \displaystyle \mu_V \left[ (2-x_1-x_2)^2+8\mu_V
  \right],\, \nonumber \\
  f_1 & = & \displaystyle x_1^2 \left( \mu_V -1+x_1\right)^2-4\mu_H
  \left(1-x_1\right)\left(1-x_1+\mu_V-\mu_V
    x_1-4\mu_V\right)\nonumber\\
  & & \displaystyle +\mu_V\left(\mu_V-4\mu_H\right)
  \left(1-4\mu_H\right)-\mu_V^2 ,\, \nonumber \\
  f_2 & = & \displaystyle
  \left(2\mu_V+x_1+x_2\right)\left[\mu_V\left(x_1+x_2-1+\mu_V-8\mu_H\right)
  \right.\nonumber \\
  & & \displaystyle
  \left. -\left(1-x_1\right)\left(1-x_2\right)\left(1+\mu_V\right)\right]+\left(1-x_1\right)^2
  \left(1-x_2\right)^2 \nonumber \\
  & & \displaystyle
  +\left(1-x_1\right)\left(1-x_2\right)\left[\mu_V^2+1+4\mu_H\left(1+\mu_V\right)\right]
  \nonumber \\
  & & \displaystyle +4\mu_H\mu_V\left(1+\mu_V+4\mu_H\right)+\mu_V^2
  ,\,  \nonumber \\
  f_3 & = & \displaystyle x_1
  \left(x_1-1\right)\left(\mu_V+x_1-1\right)-\left(1-x_2\right)\left(2-x_1\right)\left(1-x_1+\mu_V\right)
  \nonumber \\
  & & \displaystyle +2\mu_V\left(\mu_V+1-4\mu_H\right) ,\, \nonumber \\
  & & \nonumber \\
  C_{HHH} & = &
  \frac{v}{M_V^2}\, \frac{\lambda_{HHH}}{x_1+x_2-1+\mu_V-\mu_H}+\frac{2}{1-x_1+\mu_H-\mu_V}
  \nonumber \\
  & & + \frac{2}{1-x_2+\mu_H-\mu_V} + \frac{1}{\mu_V} \; . 
\end{eqnarray}
The coefficient $C_{HHH}$ includes the trilinear Higgs coupling
$\lambda_{HHH}$.

In order to obtain the full hadronic section, the differential
partonic cross section of Eq.~(\ref{vhh-partonic}) is convoluted with 
the quark parton distribution functions, $f_{q}, f_{q'}$ taken at a
typical scale $\mu_F$ specified below: 
\begin{equation}
\sigma(pp\to VHH) = \sum_{q, q'} \int_{\tau_0}^1 d\tau \int_{\tau}^1 \frac{dx}{x}
f_q(x; \mu_F^2)\, f_{\bar q/ \bar q'} \left(\frac{\tau}{x};\, \mu_F^2
\right) \hat{\sigma}_{VHH}(\hat{s}=\tau s)\, ,
\end{equation}
where $s$ stands for the hadronic c.m. energy and $\tau_0=
(2M_H+M_V)^2/s$. The total partonic cross section $\hat{\sigma}_{VHH}$
has been obtained after the integration of Eq.(\ref{vhh-partonic}) over
$x_1, x_2$.

\begin{figure}[h!]
\begin{center}
\includegraphics[scale=0.75]{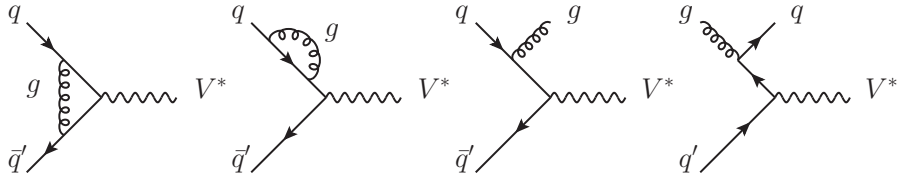}
\end{center}
\it{\vspace{-6mm}\caption{Feynman diagrams contributing to the NLO QCD
    corrections for Drell--Yan production.\label{qqV-NLO}}}
\end{figure}

The calculation of the NLO QCD corrections is similar to the
single Higgs--strahlung case. In fact, this process can be viewed as
the Drell-Yan production $pp\to V^*$ followed by the splitting process
$V^* \to VHH$. The off--shell vector boson can have any momentum $k^2$
with $(2 M_H+M_V)^2 \leq k^2 \leq \hat{s}$. This factorization is in
principle valid at all orders for the Drell--Yan like contributions
and leads, after folding with the PDF, to
\begin{equation}
\hspace{-1mm}\sigma(pp\to VHH) =\int_{\tau_0}^1
d\tau\sum_{(ij)}\frac{d\mathcal{L}^{ij}}{d\tau} \int_{\tau_0/\tau}^{1}
dz\, \hat{\sigma}^{\rm LO} (z \tau s) \Delta_{ij} (ij\rightarrow
V^*)\; , \label{eq:hovhh}
\end{equation}
with
\begin{equation}
\frac{d\mathcal{L}^{ij}}{d\tau}=
  \int_{\tau}^{1} f_i(x;\, \mu_F^2)f_{j} 
 \left(\frac{\tau}{x};\, \mu_F^2\right)
   \frac{dx}{x}\, . 
\end{equation}
In the expressions above $ij$ stands for any initial partonic
subprocess, $\Delta_{ij}$ is the Drell--Yan correction,
$z=k^2/\hat{s}$ and $\hat{\sigma}^{\rm LO}$ stands for the LO
partonic cross section of the process $q\bar{q}' \to V^* \to V HH$. At LO we have
$\displaystyle \Delta_{ij}^{\rm LO} = \delta_{i q} \delta_{j \bar
  q/\bar q'} \delta(1-z)$. In Fig.~\ref{qqV-NLO} the generic diagrams
which contribute at NLO to the Drell--Yan process $q \bar{q'}\to V^*$
are depicted. The NLO QCD corrections increase the total cross section
by $\sim +17\%$ at 14 TeV for $M_H=125$ GeV. 

\begin{figure}[h!]
\begin{center}
\includegraphics[scale=0.75]{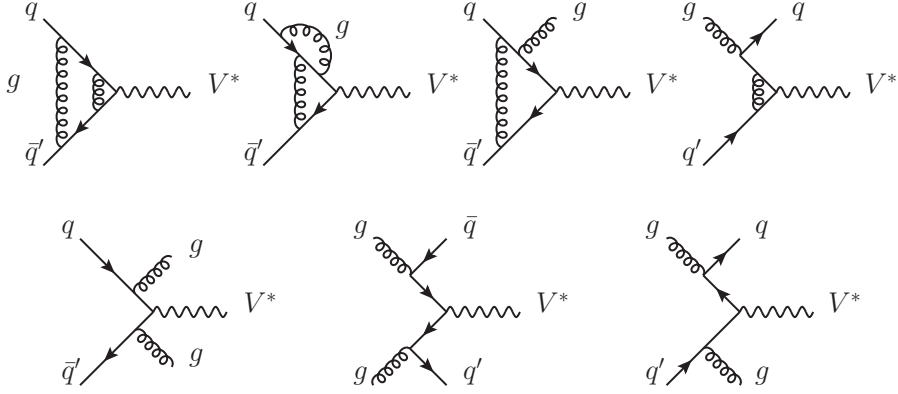}
\end{center}
\it{\vspace{-6mm}\caption{Some Feynman diagrams contributing at NNLO
    QCD to Drell--Yan production.\label{qqV-NNLO}}}
\end{figure}

We have calculated the NNLO corrections, which have not been available
so far, in the same way except for the process involving a $Z$
boson. In fact there are additional contributions that are specific to
the case of a $Z$ boson, involving an effective $Zgg$ vertex. Similar
to what is stated in Ref.~\cite{HV-NNLO} for the single Higgs
production case, only the specific gluon fusion initiated
process will be of non--negligible contribution and will be described
below. Let us start with the NNLO QCD Drell--Yan contribution. Some
generic diagrams contributing to the NNLO corrections to $q \bar q'\to
V^*$ are shown in Fig.~\ref{qqV-NNLO}.
We apply the procedure as described by Eq.~(\ref{eq:hovhh}) and the
expression is then given by
\begin{eqnarray}
  \sigma^{\rm NNLO}(pp\to VHH) & = & \sigma^{\rm LO} + \Delta
  \sigma_{q \bar q/\bar q'} + \Delta \sigma_{q g} + \Delta \sigma_{q
    q'} + \nonumber\\
  & & \Delta \sigma_{q q} +\Delta \sigma_{gg} +\delta_{VZ} \Delta
  \sigma_{gg\to ZHH},\,
\end{eqnarray}
\vspace{2mm}

with $\delta_{VZ}=1 (0)$ for $V=Z(W)$ and
\begin{eqnarray}
  \sigma^{\rm LO} & = & \sum_{q,q'} \int_{\tau_0}^1 d\tau\,
  \frac{d\mathcal{L}^{q \bar{q}/\bar{q}'}}{d\tau}\,
  \hat{\sigma}^{\rm LO}(\tau s),\, \nonumber\\
  \Delta \sigma_{q \bar q/\bar q'} & = & \sum_{q, q'}
  \left(\frac{\alpha_s(\mu_R)}{\pi}\right) \int_{\tau_0}^1 d\tau \,
  \frac{d\mathcal{L}^{q \bar{q}/\bar{q}'}}{d\tau}\, \times\nonumber\\
  & & \int_{\tau_0/\tau}^1 \hat{\sigma}^{\rm
    LO}(z\tau s) \left( \Delta^{(1)}_{q\bar q}(z)+
    \left(\frac{\alpha_s(\mu_R)}{\pi}\right) \Delta^{(2)}_{q\bar q}(z)
    \right),\, \nonumber\\
  \Delta \sigma_{q g} & = & \sum_{i = q,\bar{q}}
  \left(\frac{\alpha_s(\mu_R)}{\pi}\right) \int_{\tau_0}^1 d\tau \,
  \frac{d\mathcal{L}^{i g}}{d\tau}\, \times\nonumber\\
  & & \int_{\tau_0/\tau}^1 \hat{\sigma}^{\rm
    LO}(z\tau s) \left( \Delta^{(1)}_{q
      g}(z)+\left(\frac{\alpha_s(\mu_R)}{\pi}\right) \Delta^{(2)}_{q
      g}(z) \right),\, \nonumber\\
  \Delta \sigma_{q q'} & = & \hspace{-5mm} \sum_{i =
    q,\bar{q}, j= q',\bar{q}'}
  \left(\frac{\alpha_s(\mu_R)}{\pi}\right)^2 \int_{\tau_0}^1 d\tau \,
  \frac{d\mathcal{L}^{ij}}{d\tau}\, \int_{\tau_0/\tau}^1 \hat{\sigma}^{\rm
    LO}(z\tau s) \Delta^{(2)}_{q q'}(z),\, \nonumber\\
  \Delta \sigma_{q q} & = & \sum_{i = q,\bar{q}}
  \left(\frac{\alpha_s(\mu_R)}{\pi}\right)^2 \int_{\tau_0}^1 d\tau \,
  \frac{d\mathcal{L}^{i i}}{d\tau}\, \int_{\tau_0/\tau}^1 \hat{\sigma}^{\rm
    LO}(z\tau s) \Delta^{(2)}_{q q}(z),\, \nonumber\\
  \Delta \sigma_{g g} & = &
  \left(\frac{\alpha_s(\mu_R)}{\pi}\right)^2 \int_{\tau_0}^1 d\tau \,
  \frac{d\mathcal{L}^{g g}}{d\tau}\, \int_{\tau_0/\tau}^1 \hat{\sigma}^{\rm
    LO}(z\tau s) \Delta^{(2)}_{g g}(z),\, \nonumber\\
  \Delta \sigma_{gg\to ZHH} & = & \int_{\tau_0}^1 d\tau \,
  \frac{d\mathcal{L}^{g g}}{d\tau}\, \hat{\sigma}_{gg\to ZHH}(\tau
  s)\, .
\label{NNLO-DY}
\end{eqnarray}
The expressions for the coefficients $\Delta^{(i=1,2)}(z)$ refer to
the NLO and NNLO corrections, respectively. As they are too lengthy
to be reproduced here, we refer the reader to the appendix B
of Ref.~\cite{HV-DY} and to Ref.\cite{ggH-NNLO0}. The expressions
given there have to be rescaled by a factor of
$\left(\pi/\alpha_s\right)^i$, and $M\equiv \mu_F$, $R
\equiv\mu_R$. In our calculation we have included the full CKM matrix
elements in the quark luminosity as well as the initial bottom quark
contribution. We use the central scale
\begin{equation} 
\mu_0 = \mu_R=\mu_F = M_{VHH} \;,
\end{equation}
where $M_{VHH}$ denotes the invariant mass of the $VHH$ system.

The Drell--Yan NNLO QCD corrections Eq.~(\ref{NNLO-DY}) turn out to be
very small. They typically increase the cross section by a few percent
at 14 TeV.

\begin{figure}[h!]
\begin{center}
\includegraphics[scale=0.85]{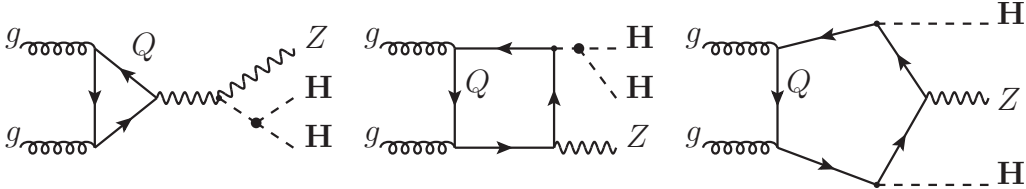}
\end{center}
\it{\vspace{-6mm}\caption{Some generic diagrams contributing to $gg\to
    ZHH$. For the triangle+box topologies, only those involving the
    trilinear Higgs couplings are depicted.\label{ggZHH-diagrams}}}
\end{figure}

The last contribution $\Delta\sigma_{gg\to ZHH}$, see diagrams in
Fig.~\ref{ggZHH-diagrams}, is only present in the case of Higgs pair
production in association with a $Z$ boson. It stems from the process
$gg\to ZHH$ which is loop--mediated already at LO. Being of order
$\alpha_s^2$ it contributes to the total cross section $pp\to ZHH$
at NNLO QCD. The process is mediated by quark loops in triangle, box
and pentagon topologies. In the latter two topologies, only top and
bottom quarks contribute as the Yukawa couplings to light quarks can
be neglected.
At the LHC the contribution of the gluon fusion channel is
substantial in contrast to the single Higgs production
case. Indeed, while in the latter the contribution is of order $\sim
+5\%$ compared to the NNLO QCD Drell--Yan contribution, in the case of 
Higgs pair production it contributes with $\sim +20\dots +30\%$
depending on the c.m. energy. This enhancement can be
explained by the additional pentagon topology which a) involves two
top Yukawa couplings and b) softens the destructive interference
between the triangle and box diagrams that is present in the single
Higgs production case. Furthermore, the suppression by a power
$(\alpha_s/\pi)^2$ is partly compensated by the increased gluon
luminosity at high energies. This explains why this channel, which has
been calculated using FeynArts/FormCalc~\cite{FormCalc}, should be
taken into account. It also implies that the scale variation in $pp\to
ZHH$ will be worse than in $pp\to WHH$ because of the ${\cal
  O}(\alpha_s^2)$ gluon fusion mechanism appearing at NNLO.

\section{Cross Sections and sensitivity at the LHC} 

\begin{figure}[h!]
\begin{center}
\includegraphics[scale=0.9]{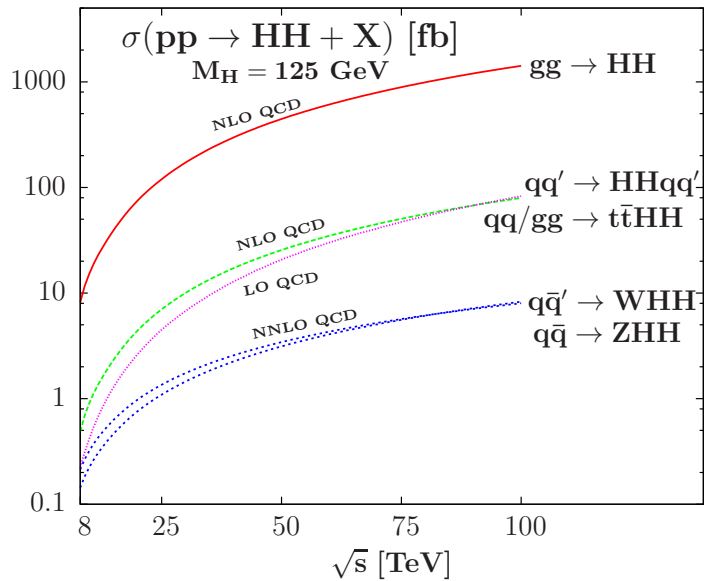}
\end{center}
\it{\vspace{-6mm}\caption{The total cross sections for Higgs pair
    production at the LHC, including
    higher-order corrections, in the main channels -- gluon fusion (red/full), VBF
    (green/dashed), Higgs-strahlung (blue/dotted), associated
    production with $t\bar t$ (violet/dotted with small dots)
    -- as a function of the c.m. energy with $M_H=125$
    GeV. The MSTW2008 PDF set has been used and higher--order
    corrections are included as discussed in section 2.\label{total_rates}}}
\end{figure}

In this section we will present the results for the calculation of the
total cross sections including the higher--order corrections discussed
in the previous section as well as the various related theoretical
uncertainties. We will use the MSTW2008 PDF set~\cite{Martin:2009iq}
as our reference set. We choose the following values for the $W$, $Z$
and top quark masses and for the strong coupling constant at LO, NLO
and NNLO,
\begin{equation}
\begin{matrix}
  M_W = 80.398\text{ GeV},\,\,\,\, M_Z = 91.1876\text{ GeV},\,\,\,\, M_t
  = 173.1\text{ GeV}, \nonumber\\
  \nonumber \\
  \alpha_s^{\rm LO}(M_Z^2) = 0.13939,\,\,\,\, \alpha_s^{\rm
    NLO}(M_Z^2) = 0.12018,\,\,\,\, \alpha_s^{\rm  NNLO}(M_Z^2) =
  0.11707.
\end{matrix}
\end{equation}
The electromagnetic constant $\alpha$ is calculated in the
$G_{\mu}$ scheme from the values of $M_W$ and $M_Z$ given above. For
the estimate of the residual theoretical uncertainties in the various
Higgs pair production processes we considered the following
uncertainties:
\begin{enumerate}
\item{ the scale uncertainty,
stemming from the missing higher order contributions and estimated by
varying the renormalization scale $\mu_R$ and the factorization scale
$\mu_F$ in the interval $\frac12 \mu_0 \leq \mu_R, \mu_F \leq 2
\mu_0$ with some restrictions on the ratio $\mu_R/\mu_F$ depending
on the process;}
\item{the PDF and related $\alpha_s$ errors. The PDFs are
    non--perturbative quantities fitted from the data and not
    calculated from QCD first principles. It is then compulsory to
    estimate the impact of the uncertainties on this fit and on the
    value of the strong coupling constant $\alpha_s(M_Z^2)$ which is
    also fitted together with the PDFs;}
\item{in the case of the gluon fusion process there is a third
    source of uncertainties which comes from the use of the effective
    field theory approximation to calculate the NLO QCD corrections,
    where top loops are taken into account in the infinite top mass
    approximation and bottom loops are neglected.} 
\end{enumerate}
In the following we will present results for $M_H=125$ GeV. Note that
the results for the total cross sections and uncertainties are nearly
the same for $M_H = 126$ GeV. The total cross sections at the LHC for
the four classes of Higgs pair production processes are shown in
Fig.~\ref{total_rates} as a function of the c.m. energy. For all
processes the numerical uncertainties are below the permille level and
have been ignored. The central scales which have been used are
($\mu_R=\mu_F=\mu_0$)
\begin{equation}
\mu_0^{gg\to HH} = M_{HH},\,\,\, \mu_0^{qq'\to HHqq'} =
Q_{V^*},\,\,\, \mu_0^{q\bar{q}'\to VHH} = M_{VHH},\,\,\,
\mu_0^{q\bar{q}/gg\to t\bar t HH} = M_t\!+\! \frac12 M_{HH}.
\end{equation}

As can be inferred from the figure and also seen in
Table~\ref{table:total_rates} the largest cross section is given by
the gluon fusion channel which is one order of magnitude 
larger than the vector boson fusion cross section. All processes are 
$\sim 1000$ times smaller than the corresponding single Higgs
production channels, implying that high luminosities are required to
probe the Higgs pair production channels at the LHC.

\begin{table}[!h]
  \renewcommand{\arraystretch}{1.3}
  \begin{center}
    \small
    \begin{tabular}{|c|ccccc|}\hline
      $\sqrt{s}$ [TeV] & $\sigma^{\rm NLO}_{gg\to HH}$ [fb] &
      $\sigma^{\rm NLO}_{qq'\to HHqq'}$ [fb] & $\sigma^{\rm
        NNLO}_{q\bar{q}'\to WHH}$ [fb] & $\sigma^{\rm
        NNLO}_{q\bar{q}\to ZHH}$ [fb] & $\sigma^{\rm
        LO}_{q\bar{q}/gg\to t\bar{t}HH}$ [fb] \\\hline
      $8$ & $8.16$ & $0.49$ & $0.21$ & $0.14$ & $0.21$
      \\\hline
      $14$ & $33.89$ & $2.01$ & $0.57$ & $0.42$ & $1.02$
      \\\hline
      $33$ & $207.29$ & $12.05$ & $1.99$ & $1.68$ & $7.91$
      \\\hline
      $100$ & $1417.83$ & $79.55$ & $8.00$ & $8.27$ & $77.82$
      \\\hline
\end{tabular}
\it{\caption[]{The total Higgs pair production
    cross sections in the main channels at the LHC
    (in fb) for given c.m. energies (in TeV) with
    $M_H=125$ GeV. The central scales which have been used are
    described in the text.}
  \label{table:total_rates}}
\end{center}
\vspace*{-7mm}
\end{table}

\subsection{Theoretical uncertainties in the gluon channel}

\subsubsection{Theoretical uncertainty due to missing higher order
  corrections}

The large $K$--factor for this process of about $1.5-2$ depending on
the c.m. energy shows that the inclusion of higher order corrections
is essential. An estimate on the size of the uncertainties due to the
missing higher order corrections can be obtained by a variation of the
factorization and renormalization scales of this process. In analogy
to single Higgs production studies~\cite{LHCXS,JB} we have estimated
the error due to missing higher order corrections by varying $\mu_R$,
$\mu_F$ in the interval
\begin{equation}
 \frac{1}{2}\mu_0\leq\mu_{R}=\mu_{F}\leq 2\,\mu_{0} \;.
\label{scalevar}\end{equation}

\begin{figure}[h!]
\begin{center}
\includegraphics[scale=0.8]{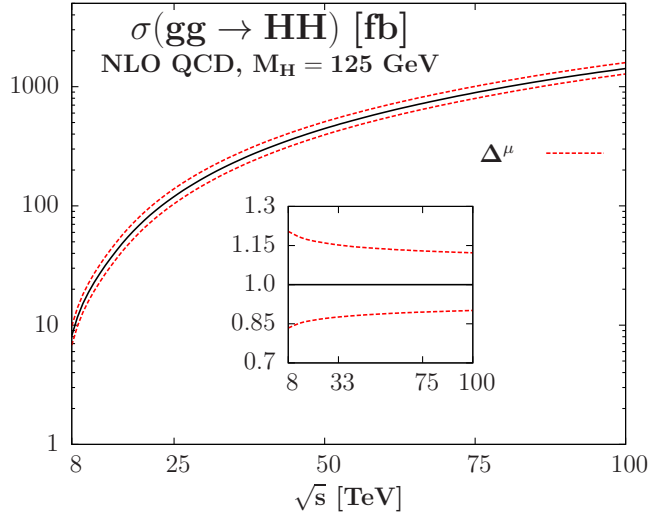}
\end{center}
\it{\vspace{-6mm}\caption{Scale uncertainty for a scale variation in
    the interval $\frac12 \mu_0 \leq \mu_R=\mu_F\leq 2\mu_0$ in
    $\sigma(gg\to HH)$ at the LHC as a function of $\sqrt{s}$ at
    $M_H=125$ GeV. In the insert the relative deviations to
    the results for the central scale $\mu_0 =
    \mu_R=\mu_F=M_{HH}$ are shown.\label{fig:ggHH_scale}}}
\end{figure}

As can be seen in Fig.~\ref{fig:ggHH_scale} we find sizeable scale
uncertainties $\Delta^{\mu}$ of order $\sim +20\%/\!-\!17\%$ at 8
TeV down to $+12\%/\!-\!10\%$ at 100 TeV. Compared to the single Higgs
production case the scale uncertainty is twice as
large~\cite{LHCXS,JB}. However, this should not be a surprise as there
are NNLO QCD corrections available for the top loop (in a heavy top
mass expansion) in the process $gg\to H$ while they are unknown for
the process $gg\to HH$.

\subsubsection{The PDF and $\alpha_S$ errors}

\begin{figure}[t!]
\begin{center}
\includegraphics[scale=0.8]{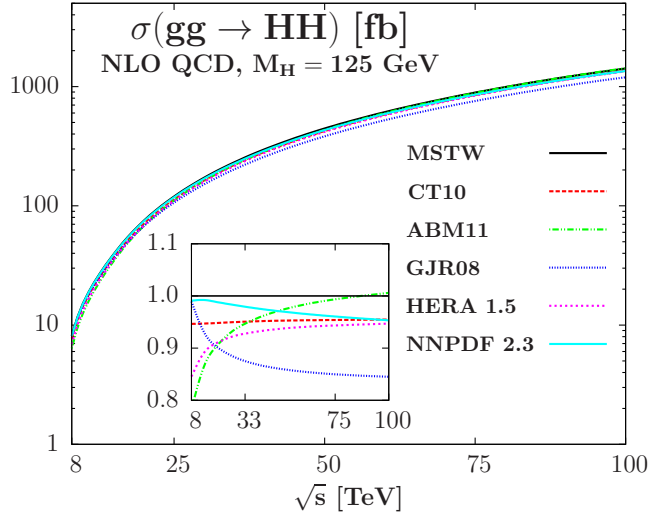}
\end{center}
\it{\vspace{-6mm}\caption{The NLO cross section
    $\sigma(gg\to HH + X)$ at the LHC as a function of the
    c.m. energy for $M_H=125$ GeV, when using different
    NLO PDF sets. In the insert the cross
    sections normalized to the cross
    section calculated with the MSTW PDF set are shown.\label{ggHH_diffpdfs}}}
\end{figure}
The parametrization of the parton distribution functions is
another source of theoretical uncertainty. First there are pure
theoretical uncertainties coming from the assumptions made on the 
parametrization, e.g. the choice of the parametrization, the set of
input parameters used, etc. Such uncertainties are rather difficult to
quantify. A possibility might be to compare different parameter
sets, such as MSTW~\cite{Martin:2009iq}, CT10~\cite{CT10},
ABM11~\cite{ABKM}, GJR08~\cite{GJR}, HERA 1.5~\cite{HERA} and NNPDF
2.3~\cite{NNPDF}.
This is exemplified in Fig.~\ref{ggHH_diffpdfs} where the predictions
using the six previous PDF sets are displayed. As can be seen there
are large discrepancies over the whole considered c.m.~energy range. At
low energies the smallest prediction comes from ABM11 which is $\sim
22\%$ smaller than the prediction made with the MSTW set while at high
energies ABM11 and MSTW lead to similar results whereas the result
obtained with the GJR08 set deviates by $\sim -15\%$. The CT10
predictions show about $-5\%$ difference over the whole energy range
with respect to the cross section obtained with MSTW while the HERA
prediction starts from lower values and eventually reaches the CT10
result. Finally the cross sections calculated with the NNPDF set 
decrease over the energy range, starting from being very similar to
the MSTW result to reach at $\sqrt{s} = 100$ TeV the one calculated
with CT10.

Another source of uncertainty due to the PDF sets comes from the
experimental uncertainties on the fitted data. The so-called
Hessian method, used by the MSTW collaboration, provides additional 
PDF sets next to the best-fit PDF. Additional $2 N_{PDF}$ sets reflect
the $\pm 1\sigma$ variation around the minimal $\chi^2$ of all
$N_{PDF}$ parameters that enter the fit. Using the $90\%$ CL error PDF
sets provided by the MSTW collaboration a PDF error of about $6\%$ is
obtained for $\sqrt{s}=8$ TeV. The uncertainty shrinks to $\sim 2\%$
for $\sqrt{s}=100$ TeV.

In addition to the PDF uncertainties, there is also an uncertainty due
to the errors on the value of the strong coupling constant
$\alpha_s$. The MSTW collaboration provides additional PDF sets such
that the combined PDF+$\alpha_s$ uncertainties can be
evaluated~\cite{MSTWalfas}. At NLO the MSTW PDF set uses
\begin{equation}
  \alpha_s(M_Z)=0.12018^{+0.00122}_{-0.00151}\text{(at 68$\%$ CL) or
  }^{+0.00317}_{-0.00386}\text{ (at 90$\%$ CL)}\;.
\label{error_alphaS_exp}
\end{equation}
As the LO process is already $\mathcal{O}(\alpha_s^2)$, uncertainties
in $\alpha_s$ can be quite substantial.
\par
The combined PDF and $\alpha_s$ error is much larger than the pure PDF
error. At 8 TeV the PDF error of $+5.8\%/\!-\!6.0\%$ rises to a combined
error of $+8.5\%/\!-\!8.3\%$. At 33 TeV the rise is even larger -- from
the pure PDF error of $+2.5\%/\!-\!2.7\%$ to the combined PDF+$\alpha_s$
error of $+6.2\%/\!-\!5.4\%$.

There is also a theoretical uncertainty on $\alpha_s$ stemming from
scale variation or ambiguities in the heavy flavour scheme
definition. The MSTW collaboration estimates this uncertainty for
$\alpha_s$ at NLO to $\Delta\alpha_s(M_Z)=\pm
0.003$~\cite{MSTWalfas}. However this uncertainty is already included
in the scale uncertainty on the input data sets used to fit the PDF
and has been taken into account in the definition of the
PDF+$\alpha_s$ uncertainty. Thus, it does not have to be taken into
account separately and the combined PDF+$\alpha_s$ error calculated
with the MSTW 2008 PDF set will be our default PDF+$\alpha_s$
uncertainty.

However, even if all these uncertainties for the MSTW PDF set are taken
into account, the different PDF set predictions do not agree. There
might be agreement if also uncertainties of the other PDF sets are
taken into account, as done in Ref.~\cite{JB} for the case of single
Higgs production. This means that the PDF uncertainty might be
underestimated, but this issue is still an open debate (see for
example Ref.~\cite{NNPDF-new} for a new discussion about theoretical
issues in the determination of PDFs) and improvements may come with
the help of new LHC data taken into account in the fits of the various
PDF collaborations.

\subsubsection{ The effective theory approach}

The last source of theoretical errors that we consider is the use of
the LET for the calculation of the NLO corrections. At LO it was found
in Ref.~\cite{pp-ggHH-LO0} that for $\sqrt{s}=16$ TeV the LET
underestimates the cross section by $\mathcal{O}(20\%)$. Furthermore
this can be even worse for different energies, not to mention the fact
that the LET approximation produces incorrect kinematic
distributions~\cite{early}. The reason is that the LET
is an expansion in $m_{top}\gg\sqrt{\hat{s}}$. Such an expansion works
very well for single Higgs production, since $\sqrt{\hat{s}}=M_{H}$
(at LO) for the production of an on--shell Higgs boson whereas in Higgs 
pair production we have
 \begin{equation}
2 M_{H}\leq\sqrt{\hat{s}}\leq\sqrt{s} \;.
\end{equation}
This means that for Higgs pair production $m_{top}\gg\sqrt{\hat{s}}$
is never fulfilled for $M_H=125$~GeV so that the LET
approximation is not valid at LO~\cite{Gillioz:2012se}.

At NLO, however, the LET approximation works much better in case the
LO cross section includes the full mass dependence. The reason is that
the NLO corrections are dominated by soft and collinear gluon
effects. They factorize in the Born term and in the NLO correction
contributions, meaning that the $K$--factor is not strongly affected
from any finite mass effects. Based on the results for the single
Higgs case \cite{ggH-NLO} where the deviation between the exact and
asymptotic NLO results amounts to less than $7\%$ for $M_H < 700$ GeV,
we estimate the error from applying an effective field theory approach
for the calculation of the NLO corrections to $10\%$.

\subsubsection{Total uncertainty}

\begin{figure}[t]
\begin{center}
\includegraphics[scale=0.8]{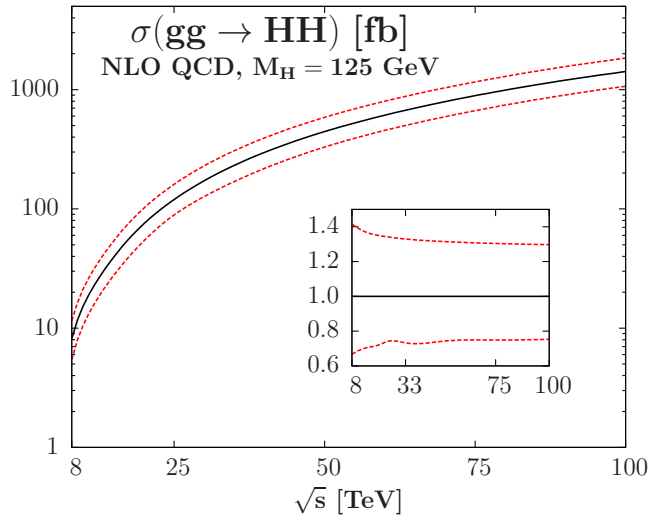}
\end{center}
\it{\vspace{-6mm}\caption{The total cross section (black/full) of the
    process $gg\to HH + X$ at the LHC for $M_H=125$ GeV as a function of
    $\sqrt{s}$ including the total theoretical uncertainty (red/dashed). The
    insert shows the relative deviation from the central cross
    section.\label{ggHH_total}}}
\end{figure}

In order to obtain the total
uncertainty we follow the procedure advocated in
Ref.~\cite{Baglio:2010um}. Since quadratic addition is too optimistic (as
stated by the LHC Higgs Cross Section Working Group, see
Ref.~\cite{LHCXS}), and as the linear uncertainty might be too
conservative, the procedure adopted is a compromise between these two
ways of combining the individual theoretical uncertainties. We first
calculate the scale uncertainty and then add on top of that the
PDF+$\alpha_s$ uncertainty calculated for the minimal and maximal
cross sections with respect to the scale variation. The LET error is
eventually added linearly. This is shown in Fig.~\ref{ggHH_total} where we display the total
cross section including the combined theoretical uncertainty. It is
found to be sizeable, ranging from $\sim +42\%/\!-\!33\%$ at 8 TeV down to
$\sim +30\%/\!-\!25\%$ at 100 TeV. The numbers can be found in
Table~\ref{table:ggHH_lhc}.

\begin{table}[h!]
 \renewcommand{\arraystretch}{1.3}
  \begin{center}
   \small
\begin{tabular}{|c|cccccc|}\hline
      $\sqrt{s}$ [TeV] & $\sigma^{\rm NLO}_{gg\to HH}$ [fb] & Scale
      [\%] & PDF [\%] & PDF+$\alpha_s$ [\%] &
      EFT [\%] & Total [\%] \\\hline
      $8$ & $8.16$ & ${+20.4}\;\;\;{-16.6}$ & ${+5.8}\;\;\;{-6.0}$ &
      ${+8.5}\;\;\;{-8.3}$ & $\pm 10.0$ & ${+41.5}\;\;\;{-33.3}$
      \\\hline
      $14$ & $33.89$ & ${+18.2}\;\;\;{-14.7}$ & ${+3.9}\;\;\;{-4.0}$ &
      ${+7.0}\;\;\;{-6.2}$ & $\pm 10.0$ & ${+37.2}\;\;\;{-29.8}$
      \\\hline
      $33$ & $207.29$ & ${+15.2}\;\;\;{-12.4}$ & ${+2.5}\;\;\;{-2.7}$ &
      ${+6.2}\;\;\;{-5.4}$ & $\pm 10.0$ & ${+33.0}\;\;\;{-26.7}$
      \\\hline
      $100$ & $1417.83$ & ${+12.2}\;\;\;{-9.9}$ & ${+2.0}\;\;\;{-2.7}$ &
      ${+6.2}\;\;\;{-5.7}$ & $\pm 10.0$ & ${+29.7}\;\;\;{-24.7}$
      \\\hline
\end{tabular}
\it{\caption[]{The total Higgs pair production
    cross section at NLO in the gluon fusion process at the LHC
    (in fb) for given c.m. energies (in TeV) at the
    central scale $\mu_F=\mu_R=M_{HH}$, for $M_H=125$ GeV. 
    The corresponding shifts due to the theoretical
    uncertainties from the various sources discussed are shown
    as well as the total uncertainty when all errors
    are added as described in the text.}
  \label{table:ggHH_lhc}}
\end{center}
\vspace*{-0.3cm}
\end{table}

\subsection{VBF and Higgs--strahlung processes}

We will not repeat the detailed description of the previous
uncertainties in this subsection and only summarize how they affect
the VBF and Higgs--strahlung inclusive cross sections. In both
channels, only the scale uncertainties and the PDF+$\alpha_s$ errors
are taken into account, the calculation being exact at a given order.

\subsubsection{The VBF channel}

\begin{figure}[h!]
\begin{center}
\begin{tabular}{cc}
\includegraphics[scale=0.7]{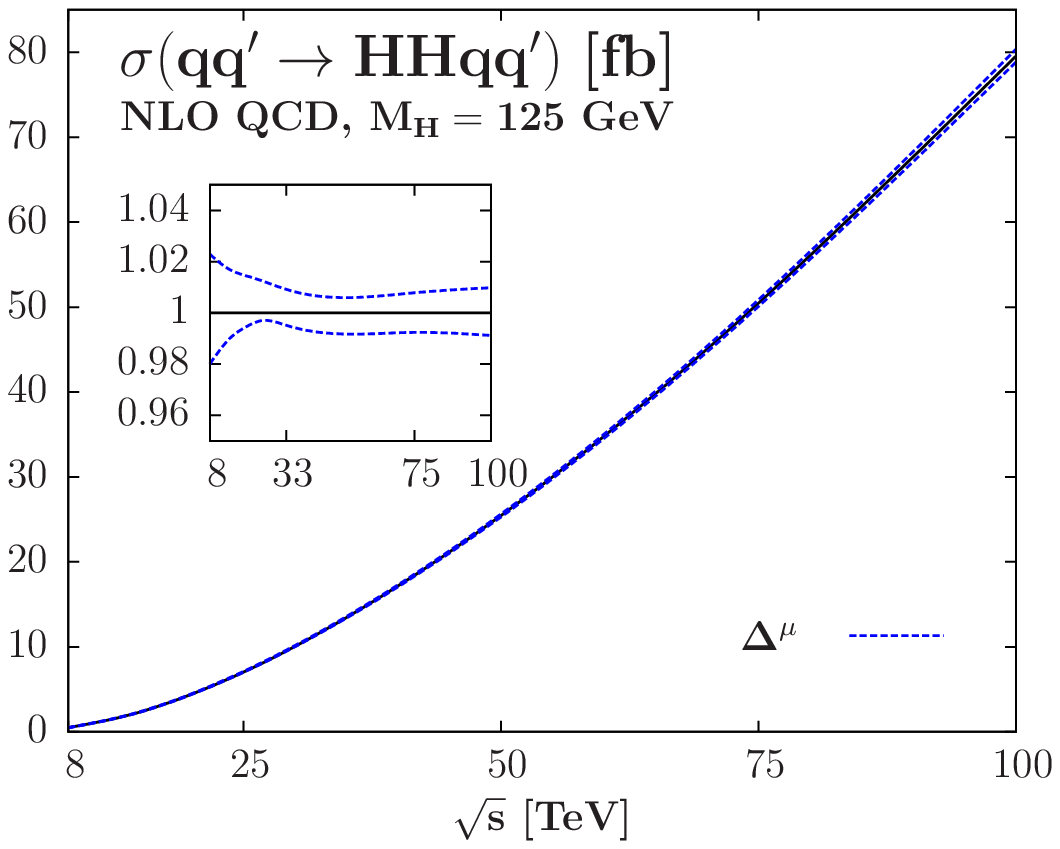}
&
\includegraphics[scale=0.7]{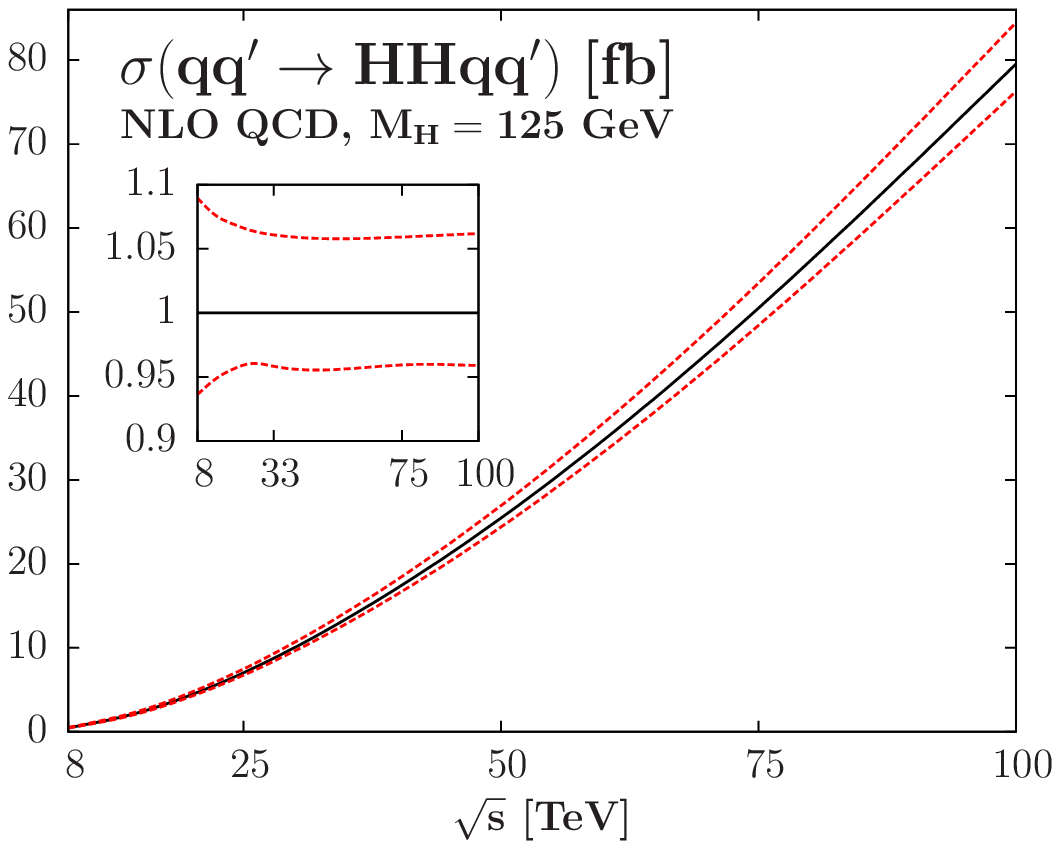}
\end{tabular}
\end{center}
\it{\vspace{-6mm}\caption{Scale uncertainty for a scale variation in
    the interval $\frac12 \mu_0 \leq \mu_R,\mu_F\leq 2\mu_0$  (left)
    and total uncertainty bands (right) in $\sigma(qq'\to HHqq')$ at the
    LHC as a function of $\sqrt{s}$ at $M_H=125$ GeV. The inserts show
    the relative deviations to the cross section evaluated at the
    central scale $\mu_0 = \mu_R=\mu_F=Q_{V^*}$.\label{VVHH_graphs}}}
\end{figure}

As stated in section 2 we use the central scale $\mu_0 = \mu_R = \mu_F
= Q_{V^*}$, that is the momentum transfer of the exchanged weak
boson. Note that a cut of $Q_{V^*} \geq 2$ GeV has to be applied as
stated in the previous section. The scale uncertainty is calculated in
exactly the same way as for the gluon fusion mechanism, exploring the
range $\mu_0/2 \leq \mu_R,\mu_F \leq 2 \mu_0$. We have checked that
imposing the restriction $1/2 \leq \mu_R/\mu_F \leq 2$ does not modify
the final result. We obtain very small scale uncertainties ranging
from $\sim \pm 2\%$ at 8 TeV down to $\sim \pm 1\%$ and even lower at
33 TeV as can be seen in Fig.~\ref{VVHH_graphs} (left). This is in 
sharp contrast with the $\pm 8\%$ uncertainty obtained at LO at 8 TeV
for example, which illustrates the high level of precision already
obtained with NLO QCD corrections.

The 90\% CL PDF+$\alpha_s$ uncertainties are calculated following
the recipe presented in the gluon fusion subsection. The
PDF+$\alpha_s$ uncertainty dominates the total error, ranging from
$+7\%/\!-\!4\%$ at 8 TeV down to $+5\%/\!-\!3\%$ at 100 TeV.

The total error has been obtained by adding linearly the scale and
PDF+$\alpha_s$ uncertainties, given the small variation of the cross
section with respect to the choice of the scale. This process has a
total theoretical uncertainty which is always below $10\%$, from
$+9\%/\!-\!6\%$ at 8 TeV to $+6\%/\!-\!4\%$ both at 33 and 100 TeV as
can be read off Table~\ref{table:VVHH_lhc}. The total uncertainty is
displayed in Fig.~\ref{VVHH_graphs} (right) as a function of the
c.m. energy. The QCD corrections drastically reduce the residual
theoretical uncertainty.

\begin{table}
\renewcommand{\arraystretch}{1.3}
  \begin{center}
    \small
    \begin{tabular}{|c|ccccc|}\hline
      $\sqrt{s}$ [TeV] & $\sigma^{\rm NLO}_{qq'HH}$ [fb] & Scale [\%] &
      PDF [\%] & PDF+$\alpha_s$  [\%]  & Total
      [\%] \\ \hline
      $8$ & $0.49$ & ${+2.3}\;\;\;{-2.0}$ & ${+5.2}\;\;\;{ -4.4}$ &
      ${+6.7}\;\;\;{-4.4}$ & ${+9.0}\;\;\;{-6.4}$ \\
      $14$ & $2.01$  & ${+1.7}\;\;\;{-1.1 }$ & ${+4.6}\;\;\;{-4.1}$&
      ${+5.9}\;\;\;{-4.1}$ & ${+7.6}\;\;\;{-5.1}$ \\
      $33$ & $12.05$  & ${+0.9}\;\;\;{-0.5 }$ & ${+4.0}\;\;\;{-3.7}$&
      ${+5.2}\;\;\;{-3.7}$ & ${+6.1}\;\;\;{-4.2}$ \\
      $100$ & $79.55$  & ${+1.0}\;\;\;{-0.9 }$ & ${+3.5}\;\;\;{-3.2}$&
      ${+5.2}\;\;\;{-3.2}$ & ${+6.2}\;\;\;{-4.1}$ \\ \hline
    \end{tabular}
    \it{\caption[]{The total Higgs pair production
      cross section at NLO in the vector boson fusion process at the
      LHC (in fb) for given c.m. energies (in TeV) at the
      central scale $\mu_F=\mu_R=Q_{V^*}$ for $M_H=125$ GeV. The
      corresponding shifts due to the theoretical uncertainties from
      the various sources discussed are also shown as well as the
      total uncertainty when all errors are added linearly.}
    \label{table:VVHH_lhc}}
  \end{center} 
\end{table}

\subsubsection{The associated Higgs pair production with a vector
  boson}

\begin{figure}[h!]
\begin{center}
\begin{tabular}{cc}
\includegraphics[scale=0.7]{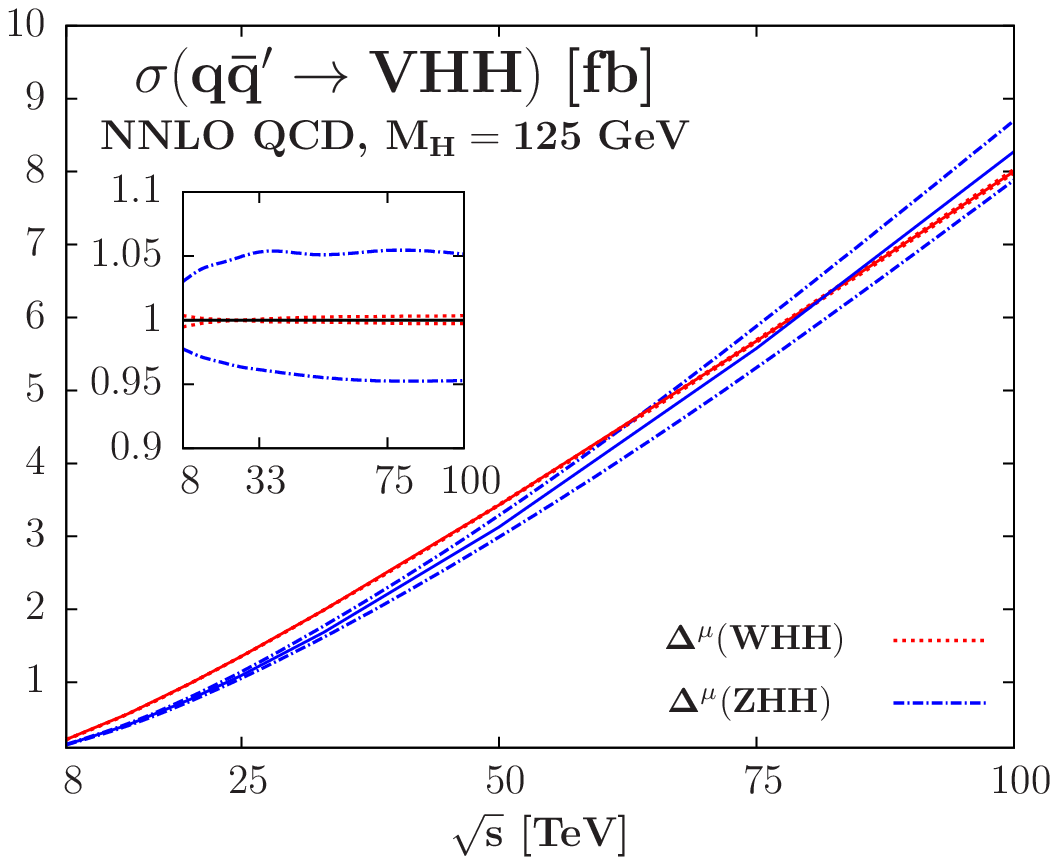}
&
\includegraphics[scale=0.7]{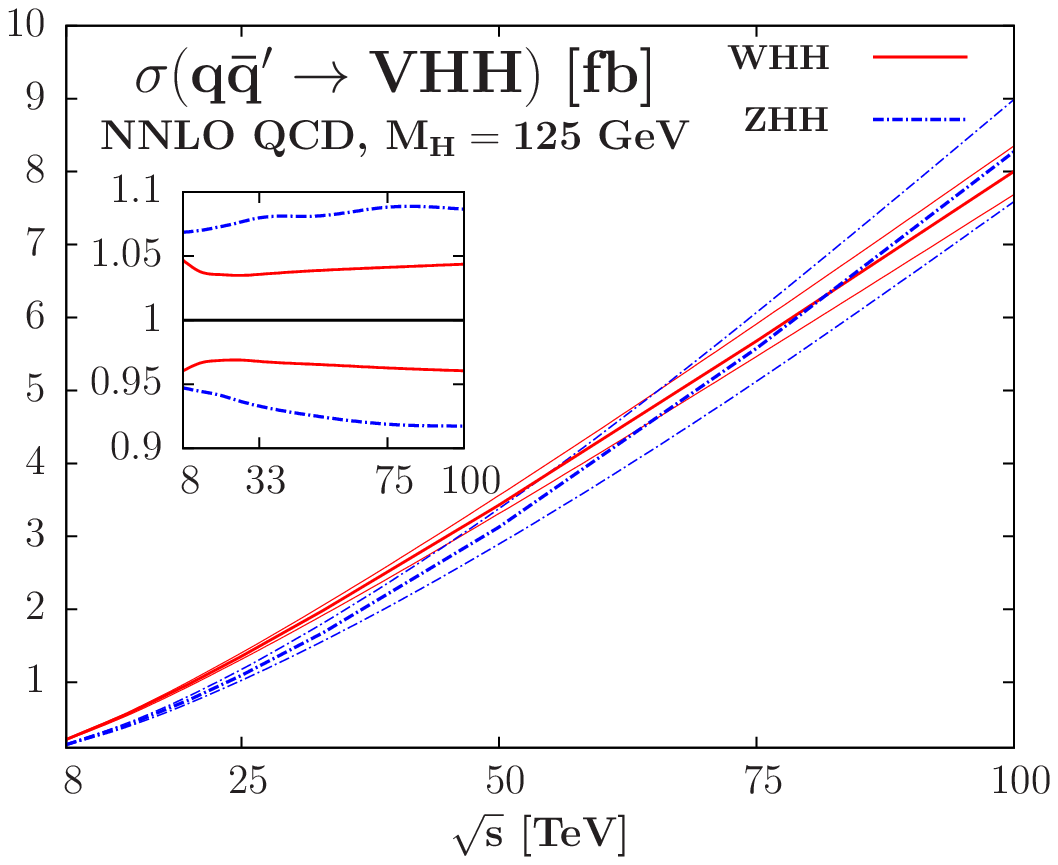}
\end{tabular}
\end{center}
\it{\vspace{-6mm}\caption{Scale uncertainty for a scale variation in
    the interval $\frac12 \mu_0 \leq \mu_R,\mu_F\leq 2\mu_0$ (left)
    and total uncertainty bands (right) in Higgs pair production
    through Higgs--strahlung at NNLO QCD at the LHC as a function of
    $\sqrt{s}$ for $M_H=125$ GeV. The inserts show the relative
    deviations to the cross section evaluated at the central scale
    $\mu_0 = \mu_R=\mu_F=M_{VHH}$.\label{VHH_graphs}}}
\end{figure}

The cross section is calculated with the central scale $\mu_0 =
\mu_R = \mu_F = M_{VHH}$ which is the invariant mass of the $W/Z$ +
Higgs pair system. The scales are varied in the interval
$\mu_0/2 \leq \mu_R = \mu_F \leq 2 \mu_0$. The factorization and
renormalization scales can be chosen to be the same as the impact of
taking them differently is totally negligible,
given the fact that the scale $\mu_R$ only appears from NLO on and
that we have a NNLO calculation which then reduces any non-negligible
contribution arising from the difference between renormalization and
factorization scales. As noticed in section 2, the scale uncertainty
is expected to be worse in the $ZHH$ channel because of the significant
impact of the gluon fusion contribution. This is indeed the case
as we obtain a scale uncertainty below $\pm 0.5\%$ in the $WHH$
channel whereas the uncertainty in the $ZHH$ channel is
$\Delta^{\mu}\sim \pm 3\%$ at 8 TeV and slightly more at higher
energies to reach $\sim \pm 5\%$ at 33 TeV, as can be seen in
Fig.~\ref{VHH_graphs} (left).

The total PDF+$\alpha_s$ error that we obtain is very similar for the
two channels $WHH$ and $ZHH$. It varies from $\sim \pm 4\%$ at 8 TeV
down to $\sim \pm 3\%$ at 33 TeV, with a slightly higher uncertainty
at 100 TeV.

The total error has been obtained exactly as in the VBF case, given
the very small variation of the cross section with respect to the
scale choice. It is dominated by the PDF+$\alpha_s$ uncertainty and
amounts to $+5\%/\!-\! 4\%$ ($+4\%/\!-\! 4\%$) at 8 (100) TeV in the $WHH$ 
channel and $+7\%/\!-\!5\%$ ($+8\%/\!-\!8\%$) at 8 (100) TeV in the $ZHH$
channel. The total theoretical uncertainty in the Higgs--strahlung
channels is less than $10\%$. The numbers are given in
Tables~\ref{table:WHH_lhc} and \ref{table:ZHH_lhc}. The total
uncertainty bands for the $WHH$ and $ZHH$ channels are displayed in
Fig.~\ref{VHH_graphs} (right).

\begin{table}
\renewcommand{\arraystretch}{1.3}
  \begin{center}
    \small
    \begin{tabular}{|c|ccccc|}\hline
      $\sqrt{s}$ [TeV] & $\sigma^{\rm NNLO}_{WHH}$ [fb] & Scale [\%] &
      PDF [\%] & PDF+$\alpha_s$  [\%]  & Total
      [\%] \\ \hline
      $8$ & $0.21$ & ${+0.4}\;\;\;{-0.5}$ & ${+4.3}\;\;\;{ -3.4}$ &
      ${+4.3}\;\;\;{-3.4}$ & ${+4.7}\;\;\;{-4.0}$ \\
      $14$ & $0.57$  & ${+0.1}\;\;\;{-0.3 }$ & ${+3.6}\;\;\;{-2.9}$&
      ${+3.6}\;\;\;{-3.0}$ & ${+3.7}\;\;\;{-3.3}$ \\
      $33$ & $1.99$  & ${+0.1}\;\;\;{-0.1 }$ & ${+2.9}\;\;\;{-2.5}$&
      ${+3.4}\;\;\;{-3.0}$ & ${+3.5}\;\;\;{-3.1}$ \\
      $100$ & $8.00$  & ${+0.3}\;\;\;{-0.3 }$ & ${+2.7}\;\;\;{-2.7}$&
      ${+3.8}\;\;\;{-3.4}$ & ${+4.2}\;\;\;{-3.7}$ \\ \hline
    \end{tabular}
    \it{\caption[]{The total Higgs pair production
      cross sections at NNLO in the $\protect{q\bar q'\to WHH}$
      process at the LHC (in fb) for different c.m.
      energies (in TeV) at the central scale $\mu_F=\mu_R=M_{WHH}$ for
      $M_H=125$ GeV. The corresponding shifts due to the theoretical
      uncertainties from the various sources discussed are also shown
      as well as the total uncertainty when all errors are added linearly.}
    \label{table:WHH_lhc}}
  \end{center} 
\end{table}

\begin{table}
\renewcommand{\arraystretch}{1.3}
  \begin{center}
    \small
    \begin{tabular}{|c|ccccc|}\hline
      $\sqrt{s}$ [TeV] & $\sigma^{\rm NNLO}_{ZHH}$ [fb] & Scale [\%] &
      PDF [\%] & PDF+$\alpha_s$  [\%]  & Total
      [\%] \\ \hline
      $8$ & $0.14$ & ${+3.0}\;\;\;{-2.2}$ & ${+3.8}\;\;\;{ -3.0}$ &
      ${+3.8}\;\;\;{-3.0}$ & ${+6.8}\;\;\;{-5.3}$ \\
      $14$ & $0.42$  & ${+4.0}\;\;\;{-2.9 }$ & ${+2.8}\;\;\;{-2.3}$&
      ${+3.0}\;\;\;{-2.6}$ & ${+7.0}\;\;\;{-5.5}$ \\
      $33$ & $1.68$  & ${+5.1}\;\;\;{-4.1 }$ & ${+1.9}\;\;\;{-1.5}$&
      ${+2.7}\;\;\;{-2.6}$ & ${+7.9}\;\;\;{-6.7}$ \\
      $100$ & $8.27$  & ${+5.2}\;\;\;{-4.7 }$ & ${+1.9}\;\;\;{-2.1}$&
      ${+3.2}\;\;\;{-3.2}$ & ${+8.4}\;\;\;{-8.0}$ \\ \hline
    \end{tabular}
    \it{\caption[]{Same as Table~\ref{table:WHH_lhc} for
        $\protect{ZHH}$ production using the central scale
        $\mu_F=\mu_R=M_{ZHH}$.}
    \label{table:ZHH_lhc}}
  \end{center} 
\end{table}

\subsection{Sensitivity to the trilinear Higgs coupling in the main
  channels}

We end this section by a brief study of the sensitivity of the three
main channels to the trilinear Higgs coupling that we want to
probe. Indeed, as can be seen in Fig.~\ref{processeslo}, all processes
do not only involve a diagram with the trilinear Higgs couplings but
also additional contributions which then dilute the sensitivity. In
order to study the sensitivity within the SM, we rescale the coupling
$\lambda_{HHH}$ in terms of the SM trilinear Higgs coupling as
$\lambda_{HHH} = \kappa \lambda_{HHH}^{\rm SM}$. This is in the same
spirit as the study done in Ref.~\cite{tripleh2} and its goal is to
give a way to estimate the precision one could expect in the
extraction of the SM trilinear Higgs coupling from $HH$ measurements
at the LHC. In particular the variation of the trilinear Higgs
coupling should not be viewed as coming from some beyond the SM
physics model and it should be noted that quite arbitrary deviations
of the trilinear Higgs couplings emerge from non-vanishing
higher-dimension operators starting with dimension 6.

\begin{figure}[h!]
\begin{center}
\begin{tabular}{cc}
\includegraphics[scale=0.65]{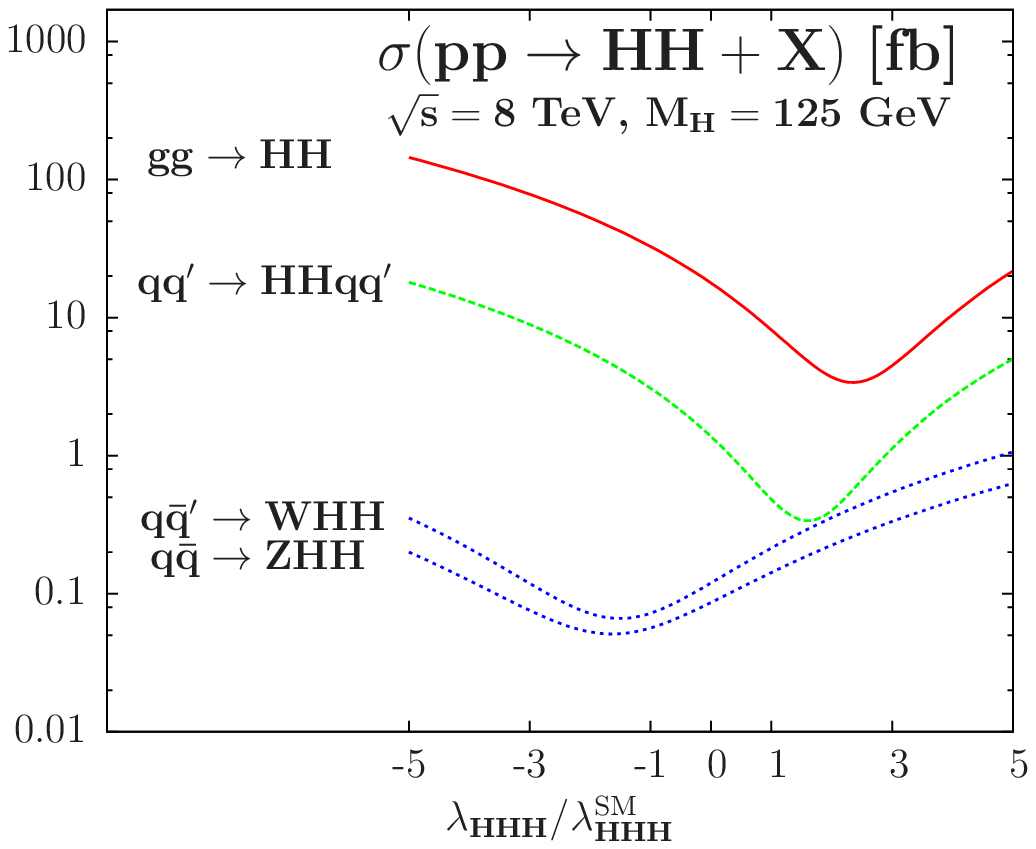}
& \includegraphics[scale=0.65]{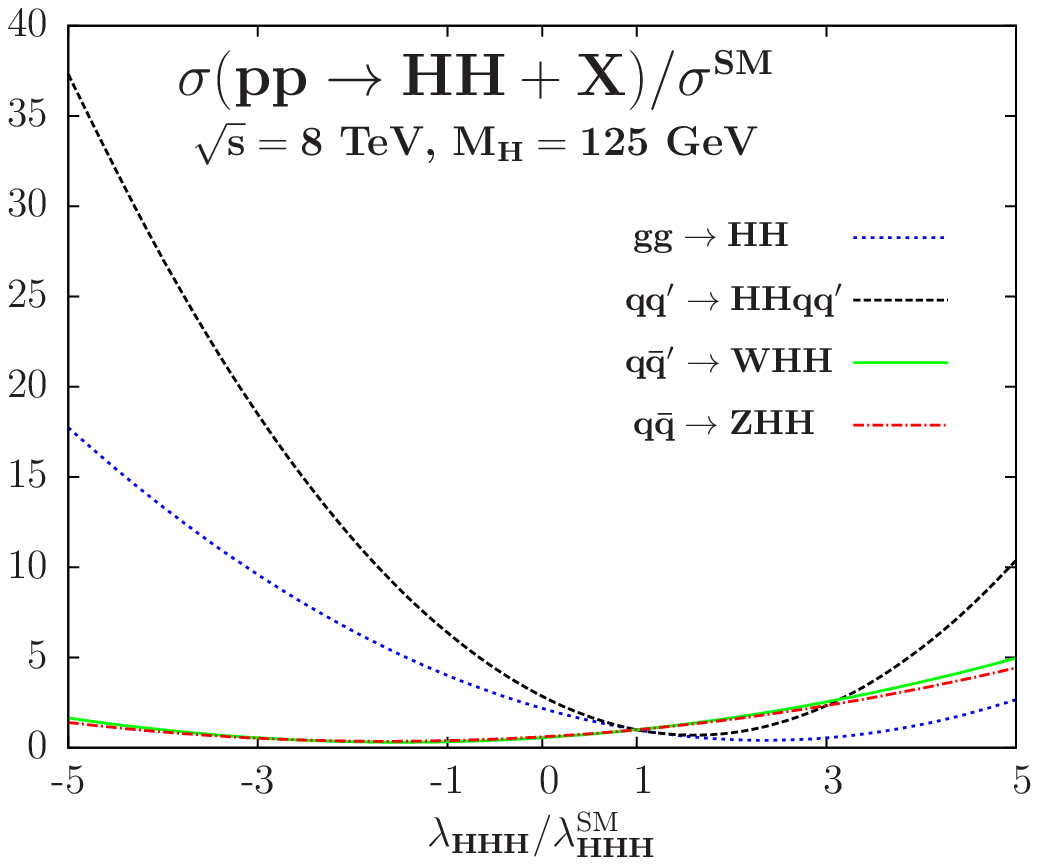}\\

& \\

\includegraphics[scale=0.65]{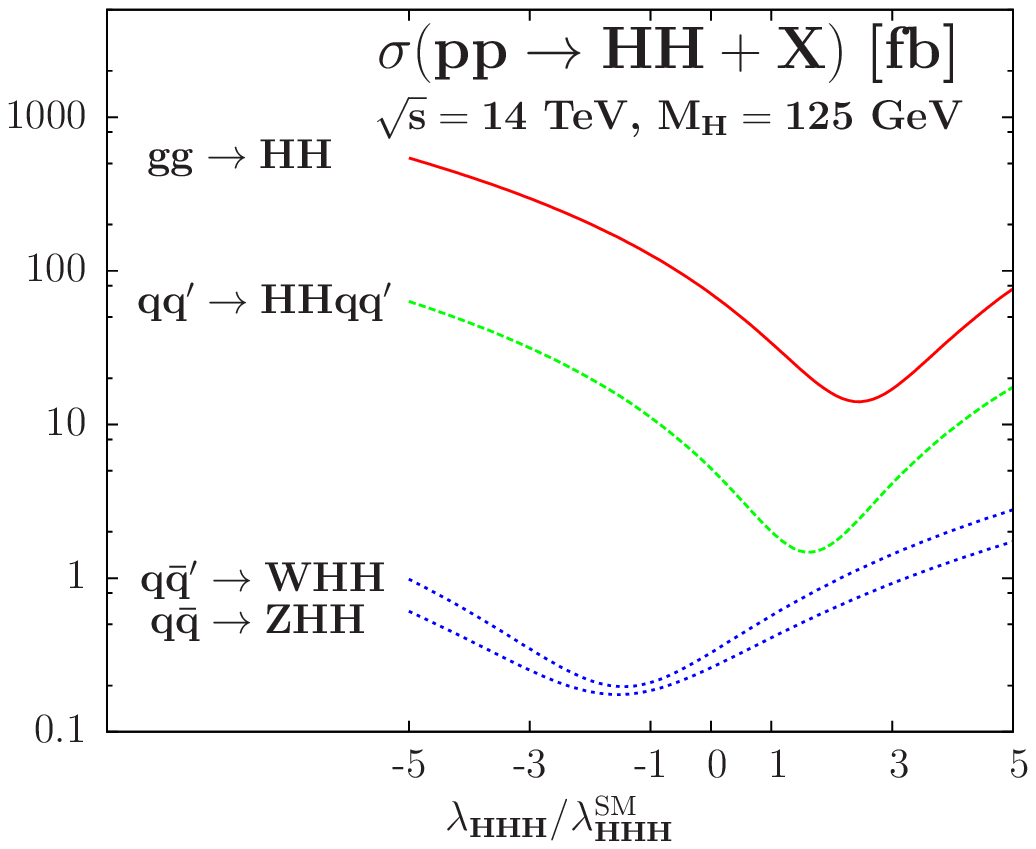}
& \includegraphics[scale=0.65]{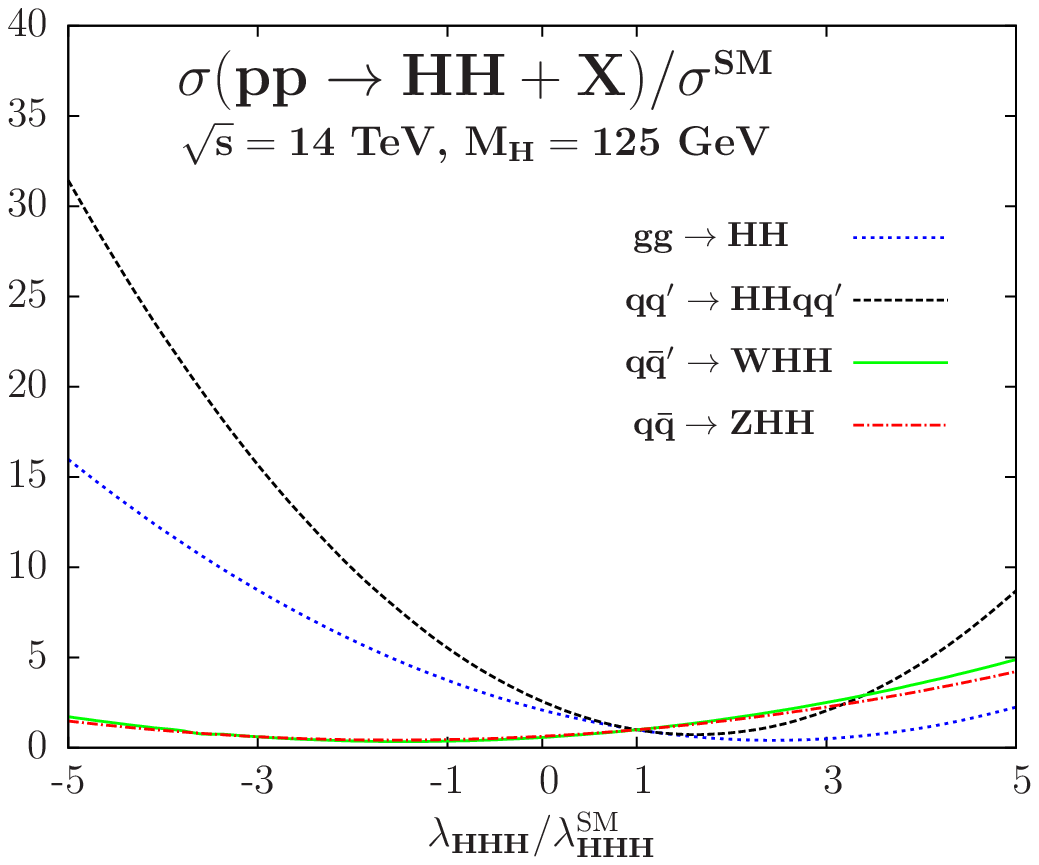}\\

& \\

\includegraphics[scale=0.65]{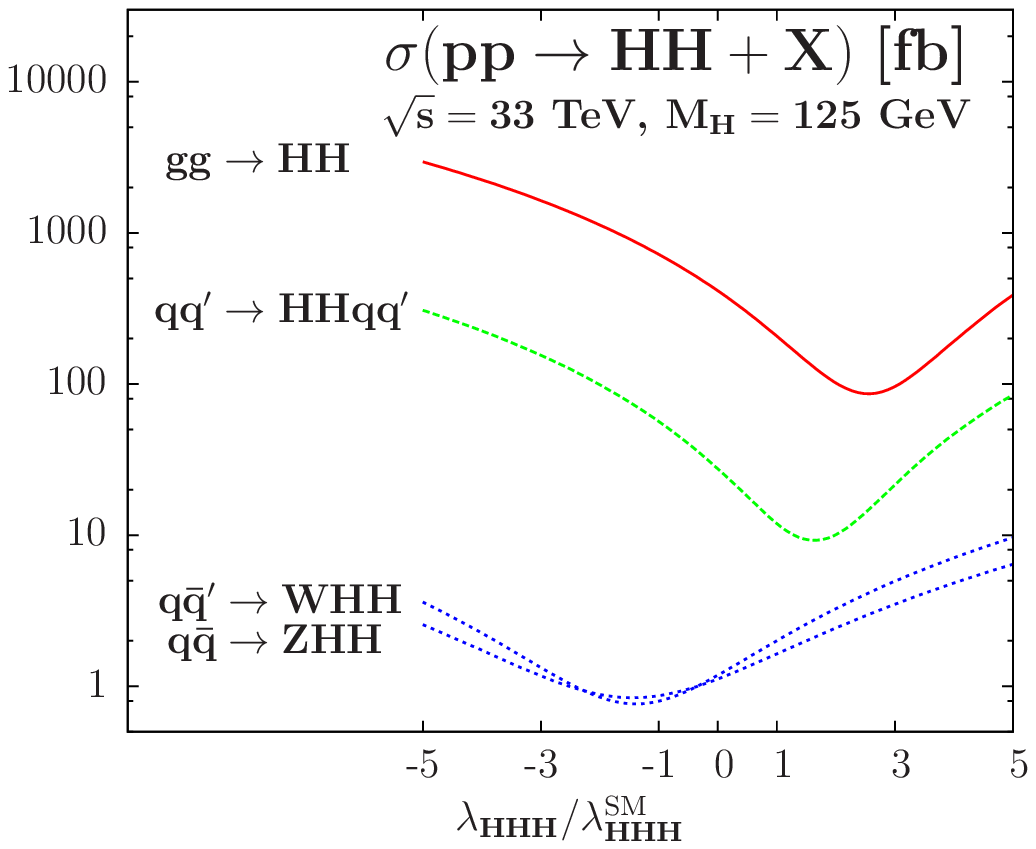}
& \includegraphics[scale=0.65]{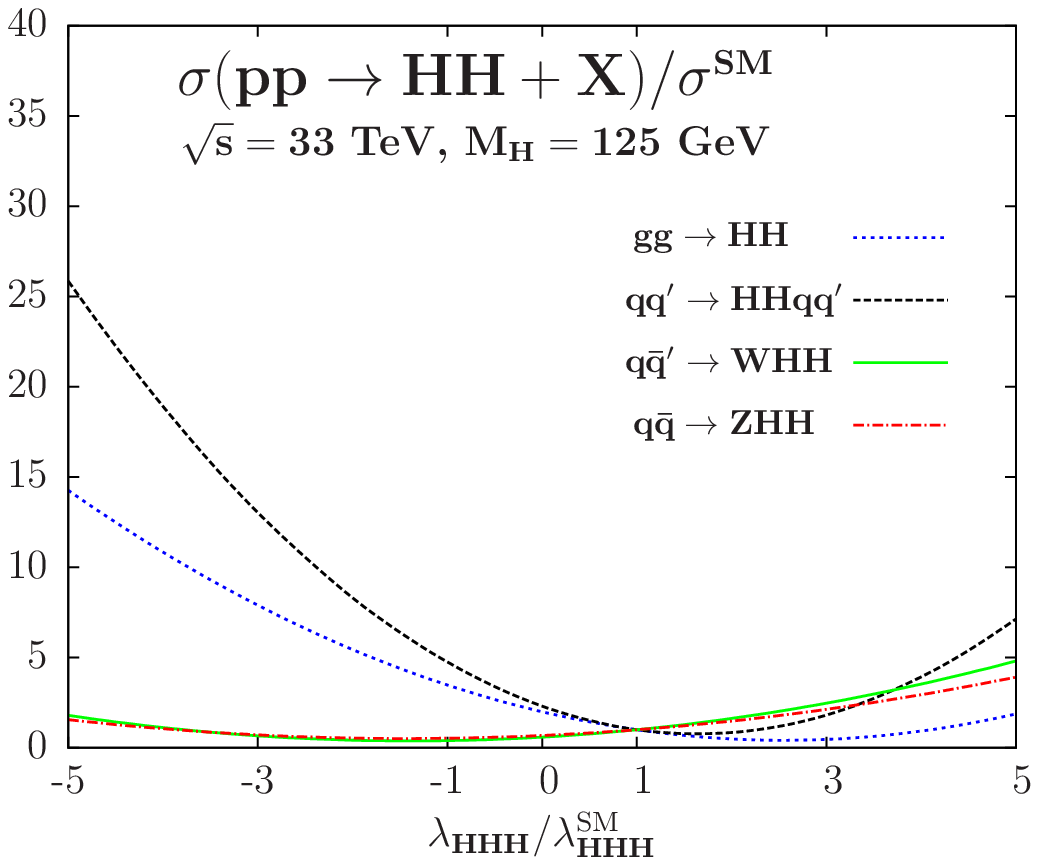}
\end{tabular}
\end{center}
\it{\caption{The sensitivity of the various Higgs pair production
    processes to the trilinear SM Higgs self--coupling at different
    c.m. energies. The left panels display the total cross sections,
    the right panels display the ratio between the cross sections at a
    given $\kappa = \lambda_{HHH}/\lambda_{HHH}^{\rm SM}$ and the
    cross sections at $\kappa=1$.\label{sensitivity}}}
\end{figure}

In Fig.~\ref{sensitivity} the three main Higgs pair production cross
sections are displayed as a function of $\kappa$ for three different
c.m. energies $\sqrt{s} = 8, 14$ and $33$ TeV. The left
panels show the total cross sections while the right panels show the
ratio between the cross sections at a given $\lambda_{HHH} = \kappa
\lambda_{HHH}^{\rm SM}$ and the SM cross section with $\kappa=1$.

As can be seen, the most sensitive channel is by far the VBF
production mode, in particular for low and high values of
$\kappa$. The shapes of the cross sections with respect to a variation
of $\kappa$ are the same in all channels and at all energies with a
minimum reached at $\kappa \sim -1$, 2 and 3 for Higgs--strahlung,
VBF and gluon fusion, respectively. The right panels of
Fig.~\ref{sensitivity} also show that the sensitivity decreases when
$\sqrt{s}$ increases. This is to be expected as the diagrams involving
the trilinear Higgs self--coupling are mediated by $s$-channel
propagators which get suppressed with increasing energy, so that the
relative importance of these diagrams with respect to the remaining
ones is suppressed.

Again it is important to note that the sensitivity
tested here does not give information on the sensitivity to Higgs
self--couplings in models beyond the SM. It only tests within the SM
how accurately the respective Higgs pair production process has to be
measured in order to extract the SM trilinear Higgs self--coupling
with a certain accuracy. For example the gluon fusion cross section
has to be measured with an accuracy of $\sim 50\%$ at $\sqrt{s}=8$~TeV
in order to be able to extract the trilinear Higgs self--coupling with
an accuracy of $50\%$, as can be inferred from Fig.~\ref{sensitivity}
upper left. Similar discussions can be found in
Refs.~\cite{tripleh2,pp-ggHH-NLO}.


\section{Prospects at the LHC} 

As shown in the previous section, inclusive Higgs boson pair
production is dominated by gluon fusion at LHC energies. Other
processes, such as weak boson fusion, $qq'\to qq'HH$, associated
production with heavy gauge bosons, $q\bar{q}'\to VHH$ ($V=W,\,Z$), or
associated production with top quark pairs, $gg/q\bar{q}\to
t\bar{t}HH$, yield cross sections which are factors of 10 --~30
smaller than that for $gg\to HH$. Since at the LHC Higgs boson pair
production cross sections are small, we concentrate on the dominant
gluon fusion process. In the following, we examine channels
where one Higgs boson decays into a $b$ quark pair and the other Higgs
boson decays into either a photon pair, $gg\to HH\to
b\bar{b}\gamma\gamma$, into a $\tau$ pair, $gg\to HH\to
b\bar{b}\tau\bar{\tau}$, or into an off--shell $W$ boson pair, $gg\to
HH\to b\bar{b}W^{*}W^{*}$. Following the Higgs Cross Section Working
Group recommendations~\cite{LHCXS}, we assume a branching ratio of
$57.7\%$ for a $125$~GeV Higgs boson decaying into $b$ quarks,
$0.228\%$ for the Higgs boson decaying into a photon pair, $6.12\%$
for the Higgs boson decaying into a $\tau$ pair and $21.50\%$ for the
Higgs boson decaying into off--shell $W^{*}$ bosons.

At the time of the analysis, no generator existed for the signal
process, but the matrix element for Higgs pair production in the 
gluon fusion channel was available in the Fortran code {\tt
  HPAIR}~\cite{pp-ggHH-NLO,hpair}. In parallel to the approach used by
the program described in~\cite{Lafaye,Blondel}, the {\tt HPAIR} matrix
element was added to  {\tt Pythia 6}~\cite{Sjostrand:2006za}. It has
been checked that the cross sections produced by {\tt HPAIR} and by
the {\tt Pythia 6} implementation are consistent. 

All tree--level background processes are calculated using {\tt
  Madgraph 5}~\cite{Alwall:2011uj}. Signal and background cross
sections are evaluated using the MSTW2008 parton distribution
functions~\cite{Martin:2009iq} with the corresponding value of
$\alpha_{s}$ at the investigated order in perturbative QCD.
The effects of QCD corrections are included in our calculation via
multiplicative factors which are summarized in the following
subsections and have been introduced in section 3 for the signal cross
sections.

\subsection{Kinematical distributions of \gghh}

\begin{figure}[!h]
  \begin{bigcenter}
    \begin{tabular}{cc}
      \includegraphics[scale=0.4]{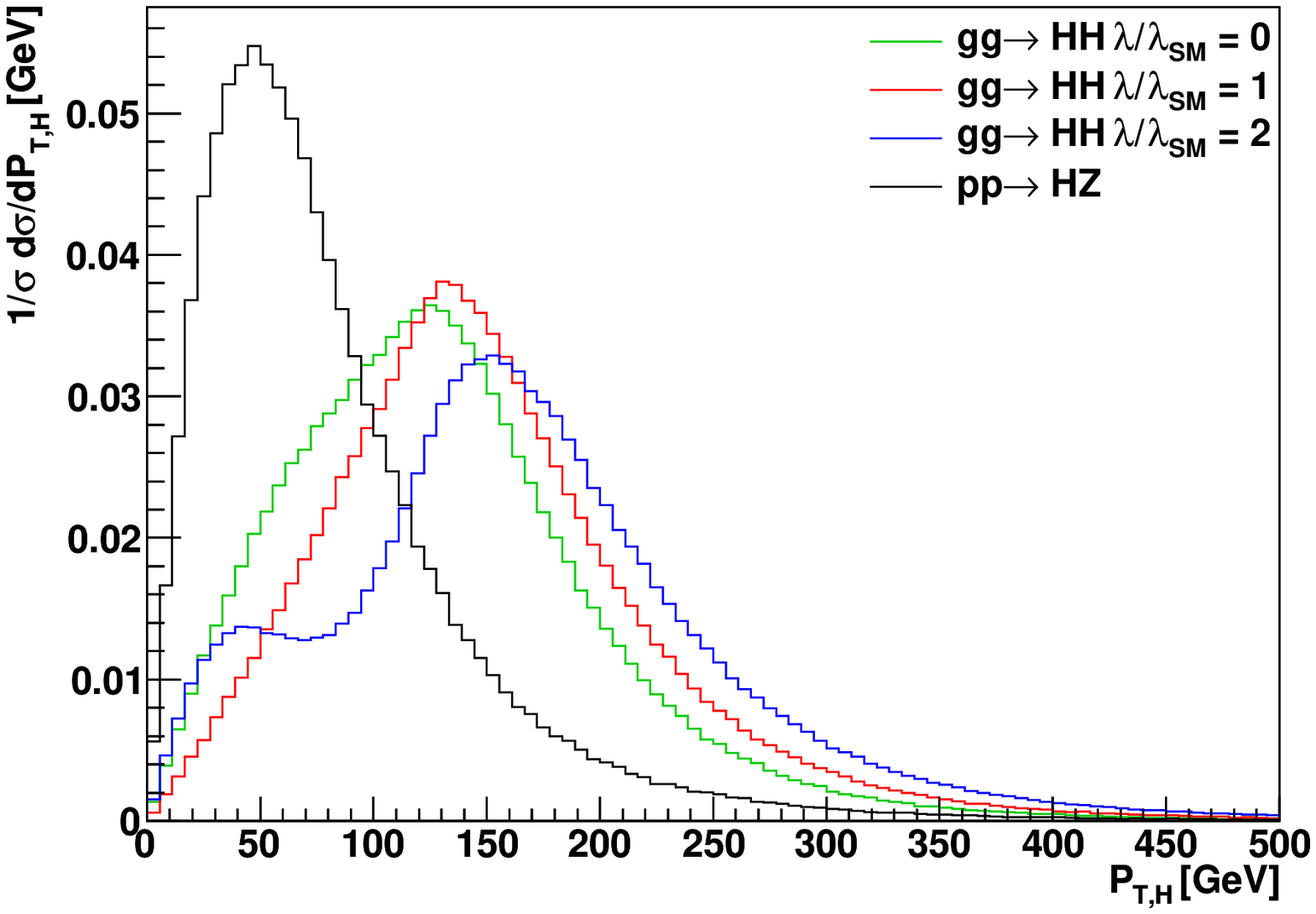}
      &
      \includegraphics[scale=0.4]{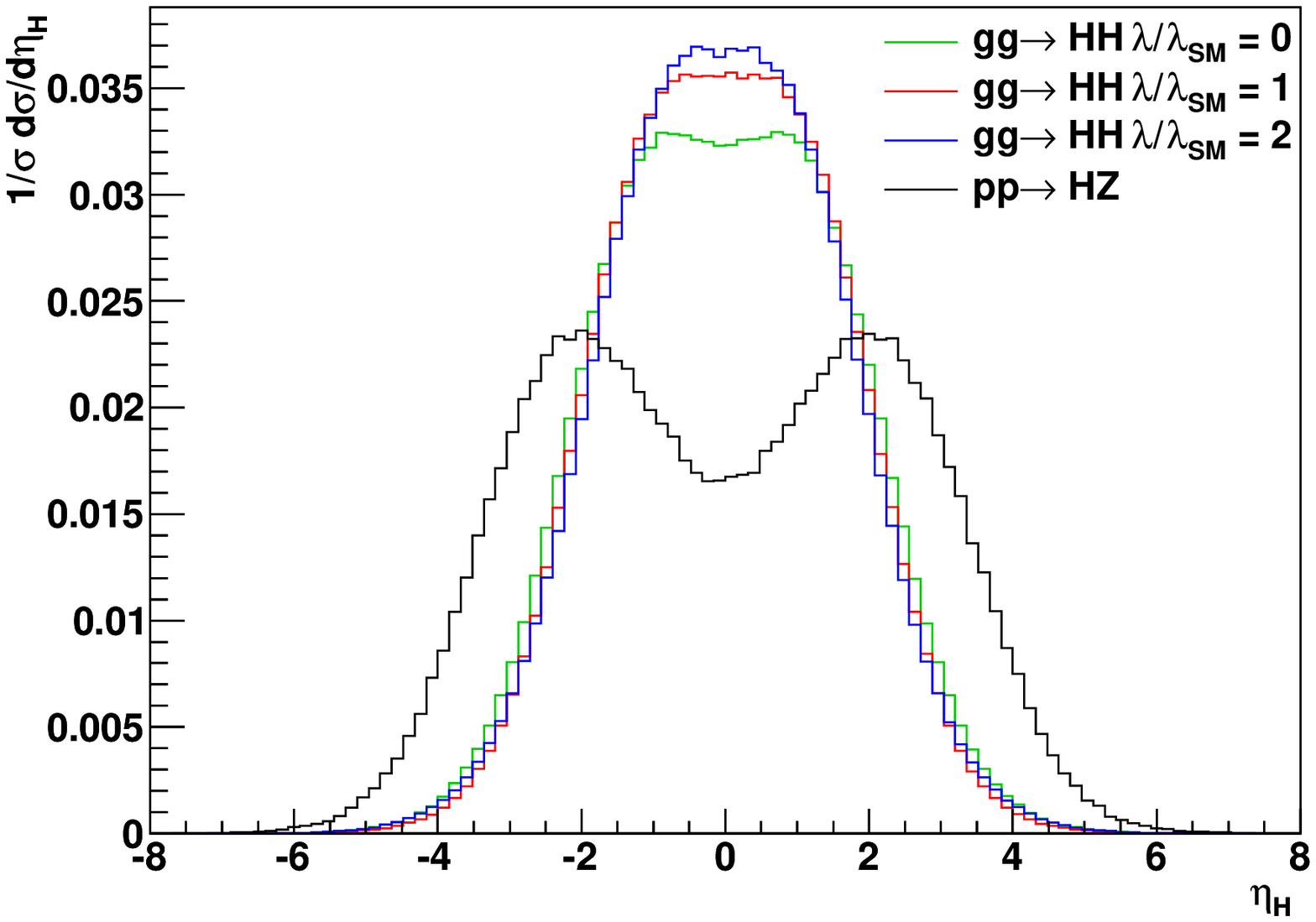}\\
    \end{tabular}
    \begin{tabular}{ccc}
      \includegraphics[scale=0.4]{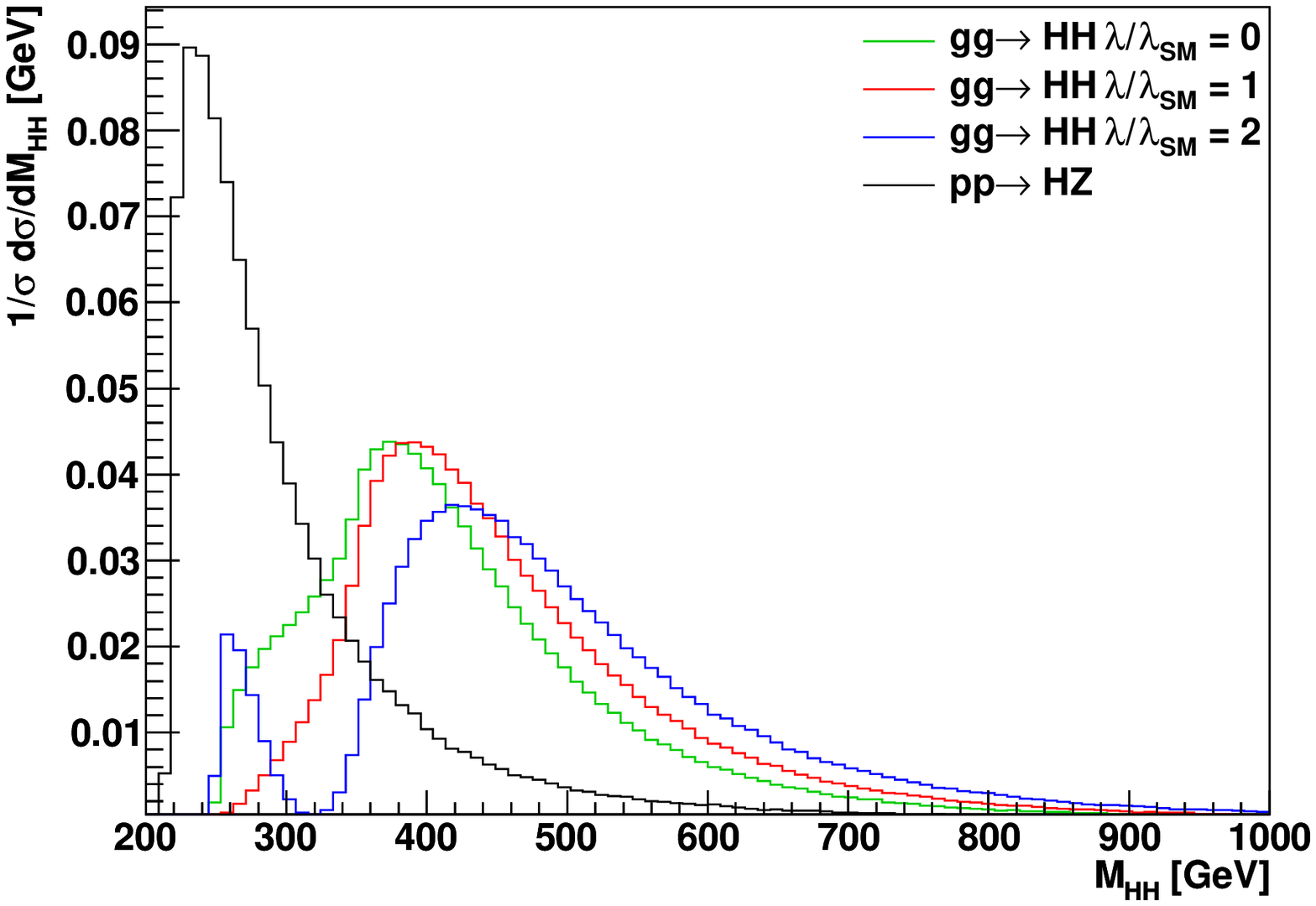}
      &
      \includegraphics[scale=0.4]{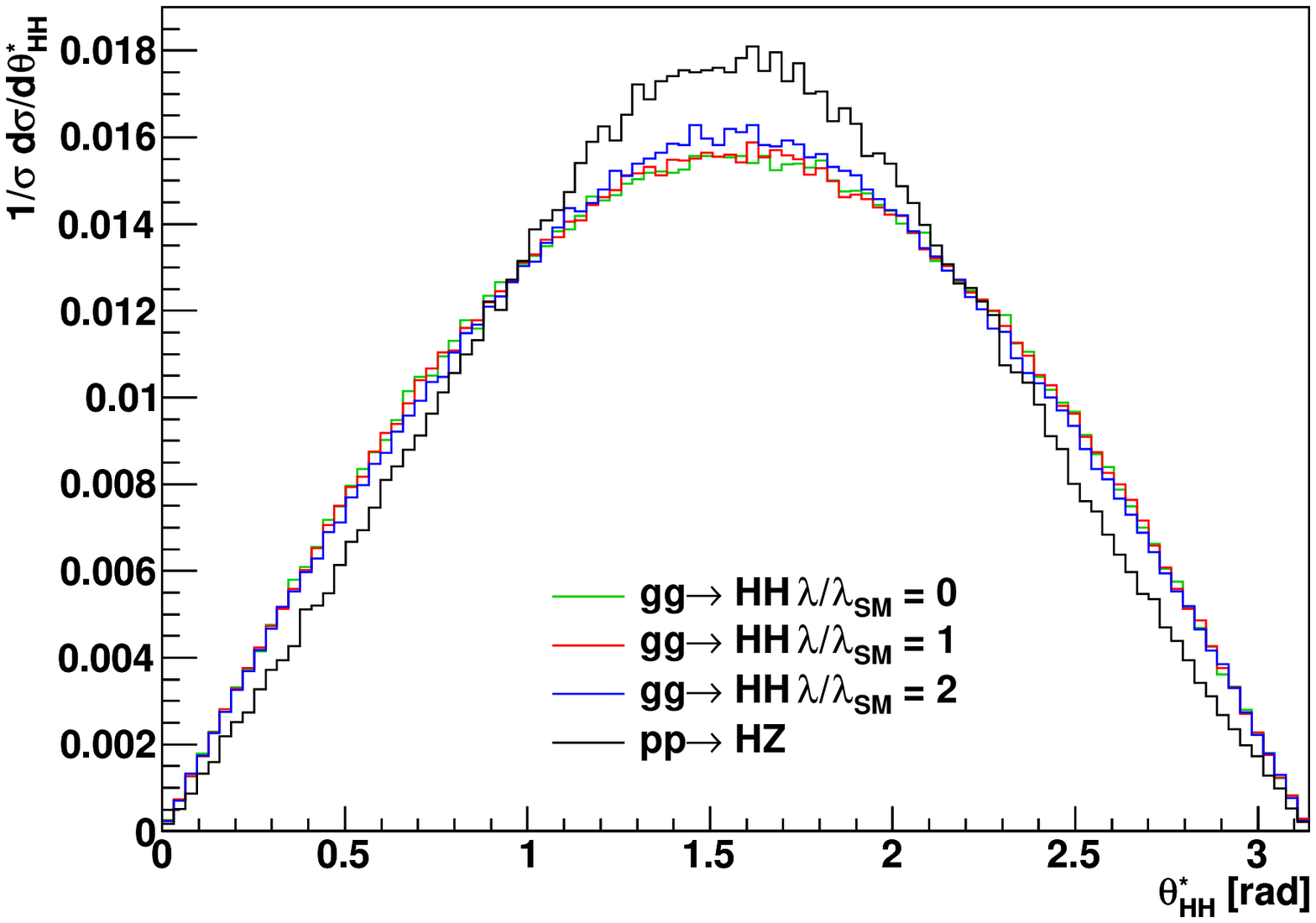}\\
    \end{tabular}
    \includegraphics[scale=0.4]{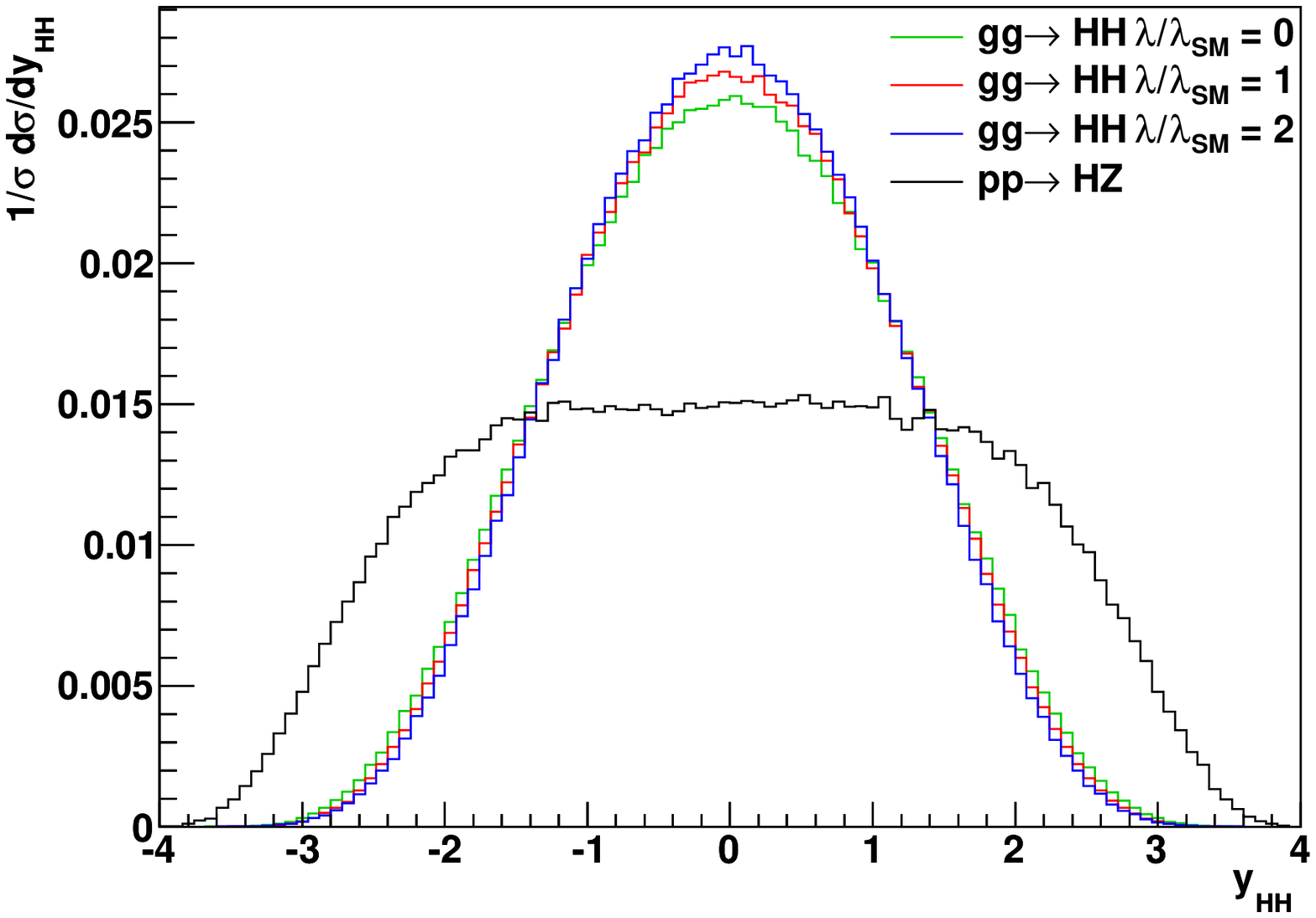}
  \end{bigcenter}
  \it{\vspace{-6mm}\caption{Normalized distributions of $\pth$, $\etah$,
      $\mhh$, $\thetas$ and $y_{HH}$ for different values of the
      trilinear Higgs coupling in terms of the SM coupling,
      $\lambda/\lambda_{SM}=0,1,2$.\label{distributions_gghh}}}
\end{figure}

In this subsection the characteristic distributions of the gluon
fusion process $\gghh$ are studied for several observables. In
Fig.~\ref{distributions_gghh}, we show for each of the two final state
Higgs bosons the normalized distributions of the transverse momentum $\pth$
and the pseudorapidity $\etah$, as well as the invariant mass $\mhh$,
the helicity angle $\thetas$ which is the angle between the off-shell
Higgs boson, boosted back into the Higgs boson pair rest frame, and the
Higgs boson pair direction, and the rapidity $y_{HH}$ of the Higgs boson
pair. The distributions of each observable are shown for different
values of the trilinear Higgs coupling $\lambda/\lambda_{SM}=0$ (green
curve), 1 (red curve) and 2 (blue curve). We also include in the plots
the typical background $q\bar{q} \rightarrow ZH$ coming from the Higgs
boson itself (black curve), the $Z$ boson faking a Higgs boson.

The Higgs bosons from inclusive Higgs pair production are typically
boosted, as we can see in the upper left plot of
Fig.~\ref{distributions_gghh} where the $\pth$ distributions reach
their maximum for $\pth \sim 150$~GeV. For $\lambda/\lambda_{SM}=2$,
the triangle diagram interferes destructively with the box diagram,
which explains the dip in the $\pth$ distribution. This high
transverse momentum spectrum can also be interpreted in terms of the
low pseudorapidity of the two Higgs bosons which have a typical
symmetric distribution with the maximum around zero, see
Fig.~\ref{distributions_gghh} upper right. The $q\bar{q} \rightarrow
ZH$ background has a completely different topology with less boosted
Higgs and $Z$ bosons, $P_{T,H/Z} \sim 50$~GeV, with pseudorapidity of
order $\vert \eta_{H/Z}\vert \sim 2$ as can be seen in the upper left
and right plots of Fig.~\ref{distributions_gghh}. The middle left plot
of Fig.~\ref{distributions_gghh} displays the distributions of the
invariant mass of the Higgs boson pair. For the signal the typical
value is $\mhh \gtrsim 400$~GeV to be compared to a lower value of
$M_{ZH} \gtrsim 250$~GeV for the $ZH$ background. Also note that an
important depletion appears in the signal for $\lambda/\lambda_{SM}=2$
caused by the destructive interference between the triangle and box
contribution. Due to the Higgs boson scalar nature a known
discriminant observable is the angle
$\thetas$~\cite{Battaglia:2001nn}. The middle right plot in
Fig.~\ref{distributions_gghh} shows that signal and $ZH$ background
have similar distributions thus making this observable less
discriminant than others described before but still efficient for some
specific backgrounds, as we will see in the following. From the bottom
plot of Fig.~\ref{distributions_gghh}, it can be inferred that the
rapidity distribution of the Higgs pair is narrower for the signal
than for the $ZH$ background.

In the following the different decay channels $HH \rightarrow \bbgg$,
$HH \rightarrow \bbtt$ and $HH \rightarrow \bbww$ will be investigated
in more detail.

\subsection{The $\bbgg$ decay channel}

In this subsection, the $\bbgg$ final state for the production of two
Higgs bosons with a mass of 125~GeV at $\sqrt{s}=14$~TeV is
investigated. Earlier studies can be found in Ref.~\cite{early}. The
calculation of the signal, $pp \rightarrow HH \rightarrow \bbgg$, is 
performed as described above by incorporating the matrix element
extracted from the program {\tt HPAIR} into {\tt Pythia 6}. We include
the effects of NLO QCD corrections on the signal by a multiplicative
factor, $K_{NLO}=1.88$, corresponding to a 125~GeV Higgs boson and a
c.m.~energy of 14~TeV. Here we set the factorization and
renormalization scales equal to $M_{HH}$. The generated background
processes are the QCD process $\bbgg$ and the associated
production of a Higgs boson with a $t\bar{t}$ pair, $t\bar{t}H$, with
the Higgs boson subsequently decaying into a photon pair and the top
quarks decaying into a $W$ boson and a $b$ quark, as well as single
Higgs-strahlung $ZH$ with the Higgs boson decaying into $\gamma\gamma$
and the $Z$ boson decaying into $b\bar b$. The QCD corrections have
been included by a multiplicative~$K$--factor applied to the
respective LO cross section. All $K$--factors are taken at NLO except
for the single Higgs--strahlung process which is taken at NNLO QCD and
the case of $b\bar{b}\gamma\gamma$ in which no higher order
corrections are taken into account. The various $K$--factors are given
in Table~\ref{K_bbgg} and taken from Ref.~\cite{LHCXS}. The
factorization and renormalization scales have been set to $M_{HH}$ for
the signal and to specific values for each process for the backgrounds.

\begin{table}
  \renewcommand{\arraystretch}{1.3}
  \begin{center}
    \small
    \begin{tabular}{c|cccc}
      $\sqrt{s}$~[TeV] & $HH$ & $b\bar{b} \gamma\gamma$ & $t\bar{t}
      H$ & $ZH$ \\ \hline
      14 & 1.88 & 1.0 & 1.10 & 1.33 
    \end{tabular}
    \it{\vspace{-2mm}\caption{$K$--factors for $gg\to HH$,
        $b\bar{b}\gamma\gamma$, $t\bar{t}H$ and $ZH$
        production at $\sqrt{s}=14$~TeV~\protect\cite{LHCXS}. The
        Higgs boson mass is assumed to be $\mh=125$~GeV.\label{K_bbgg}}}
  \end{center}
\end{table}

In this analysis, the signal and background processes are generated
with exclusive cuts. The basic acceptance cuts are motivated by the
fact that the $b$ quark pair and the photon pair reconstruct the Higgs
mass according to the resolutions expected for ATLAS and CMS. Note
that starting from this section all the plots include the decays and
the acceptance cuts specific to each final state.

In detail, we veto events with leptons having soft transverse
momentum $p_{T, \ell}>20~{\rm GeV}$ and with a pseudorapidity $\vert
\eta_{\ell} \vert<2.4$ in order to reduce the $t\bar{t} H$
background. Furthermore we also veto events with QCD jets
$p_{T,jet}>20~{\rm GeV}$ and with a pseudorapidity
$\vert\eta_{jet}\vert<2.4$ to further reduce the $t\bar{t} H$
background. We require exactly one $b$ quark pair and one photon
pair. The $b$ quark pair is restricted to have $p_{T,b}>30~{\rm GeV}$,
$\vert \eta_b \vert<2.4$ and $\Delta R(b,b) > 0.4$, where $\Delta
R(b,b)$ denotes the isolation of the two $b$ quarks defined by the
distance $\Delta R=\sqrt{(\Delta\eta)^2+(\Delta \phi)^2}$ in the
pseudorapidity and azimuthal angle plane $(\eta,\phi)$. We consider
the $b$--tagging efficiency to be $70\%$. The photon pair has to fulfill
$p_{T,\gamma}>30~{\rm GeV}$, $\vert\eta_\gamma\vert<2.4$ and $\Delta
R(\gamma,\gamma) > 0.4$. The two reconstructed Higgs bosons, from the
$b$ quark pair and from the photon pair, have to reproduce the 
Higgs boson mass within a window of 25~GeV, $112.5~{\rm GeV} \, < \,
M_{b\bar{b}} \,< \, 137.5~{\rm GeV}$, and a window of 10~GeV,
$120~{\rm GeV} \, < \, M_{\gamma\gamma} \, < \, 130~{\rm GeV}$,
respectively. We require additional isolations between the $b$ quarks
and the photons being $\Delta R(\gamma,b) > 0.4$.

\begin{figure}[!h]
  \begin{bigcenter}
    \begin{tabular}{cc}
      \includegraphics[scale=0.4]{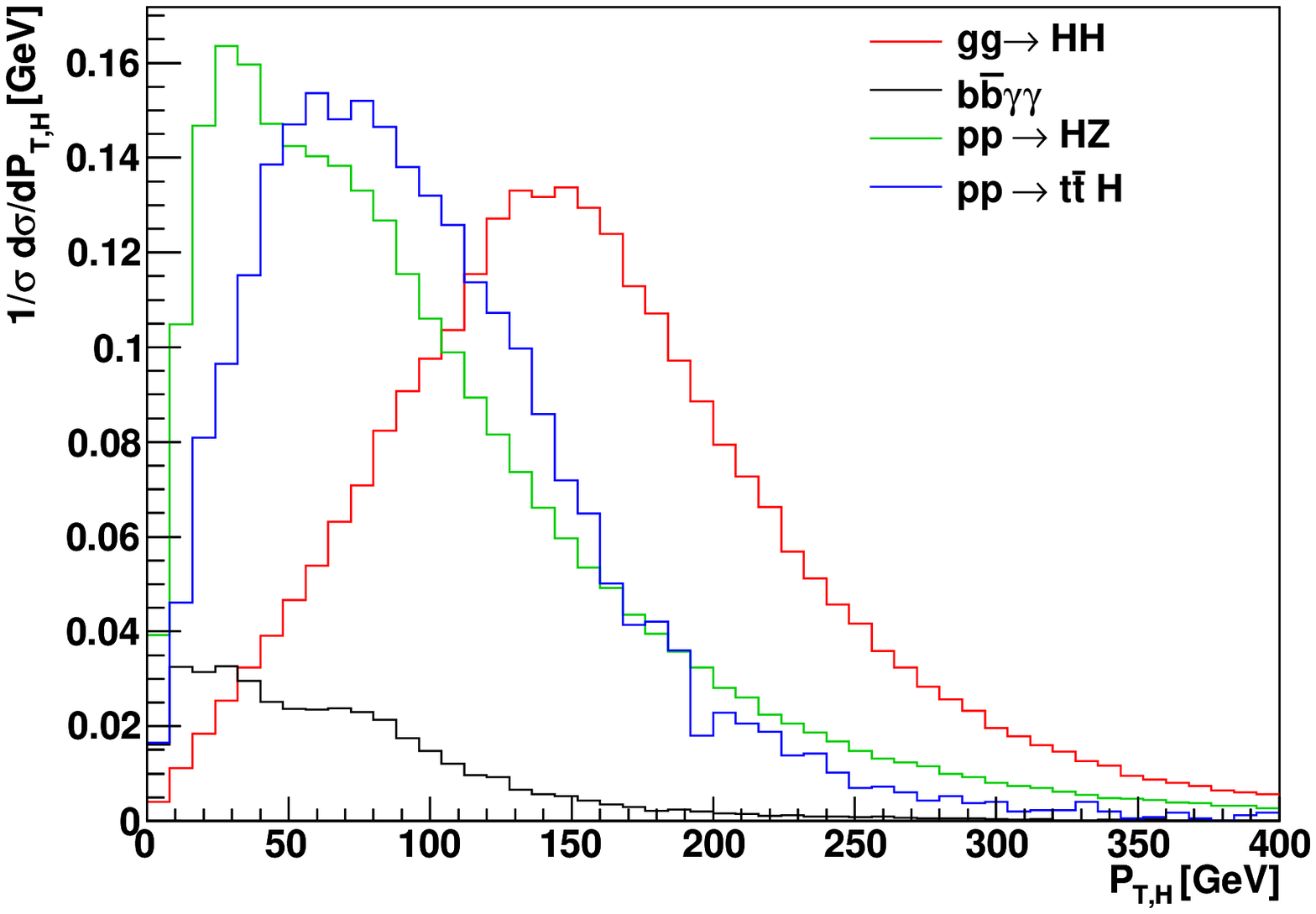}
      &
      \includegraphics[scale=0.4]{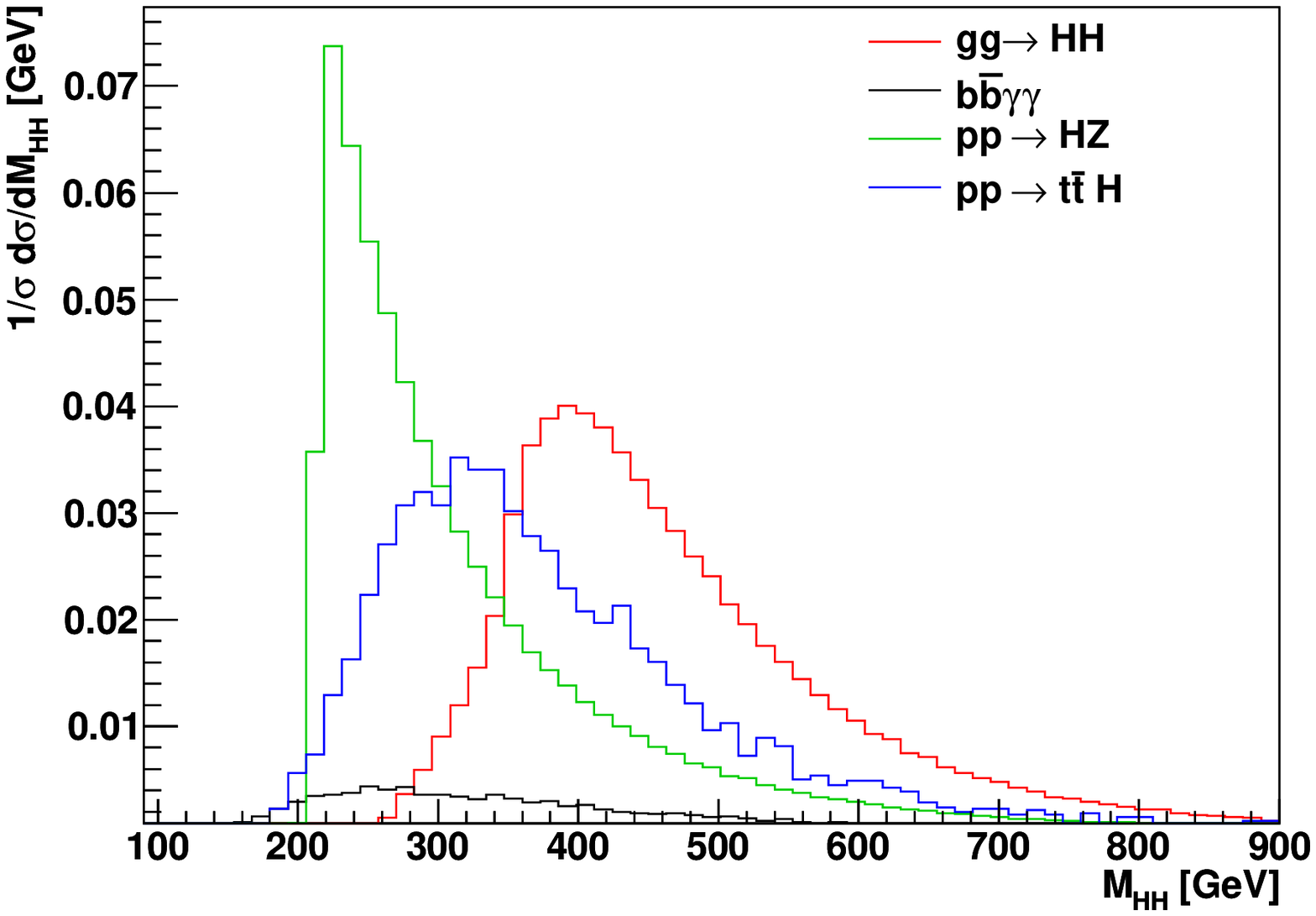}\\
    \end{tabular}
    \includegraphics[scale=0.4]{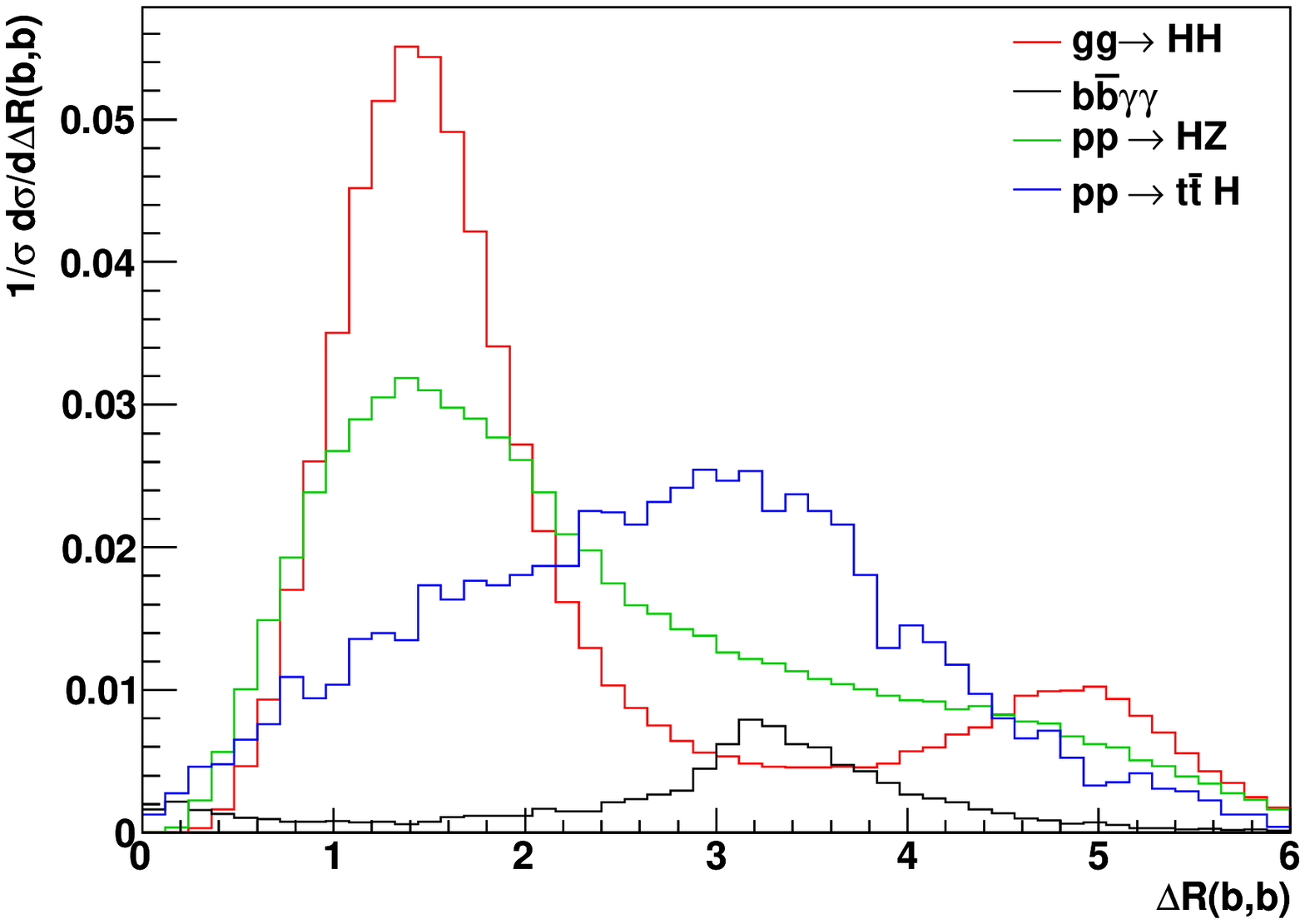}
  \end{bigcenter}
  \it{\vspace{-6mm}\caption{Normalized signal and backgrounds
      distributions of $\pth$, $\mhh$ and $R_{bb}$ in the
      $b\bar{b}\gamma\gamma$ channel.\label{distributions_bbgg}}}
\end{figure}

Based on the distributions shown in Fig.~\ref{distributions_bbgg},
apart from the acceptance cuts we have applied more advanced cuts for
this parton level analysis. We first require the reconstructed
invariant mass of the Higgs pair to fulfill $\mhh >$ 350
GeV. Furthermore we remove events which do not satisfy $\pth >$ 100
GeV. We also constrain the pseudorapidity of the two reconstructed
Higgs bosons, $|\etah| < 2$, and the isolation between the two $b$
jets to be $ \Delta R(b,b) < 2.5$.

The results are collected in Table~\ref{table_bbgg}. The local
decrease of the sensitivity between the cut on $\mhh$ and the cut on
$\pth$ is explained by the fact that we accept to have a reduced
sensitivity locally during the chain of cuts in order to enhance the
final significance. In the case described in this section a cut on
$\pth$ alone reduces the sensitivity as does a cut on $\etah$ alone, but
the first cut actually improves the discrimination between the signal
and the background in the pseudorapidity distribution, hence allowing
for a larger improvement when applying the $\etah$ cut just after the
$\pth$ cut. Eventually all the cuts allow for an improvement of the
significance by two orders of magnitude, that is the ratio of signal
events $S$ over the square root of background events $B$,
$S/\sqrt{B}$. The final value for $S/\sqrt{B}$ is $16.3$ for an
integrated luminosity of $\lum=3000$ fb$^{-1}$, corresponding to 51
signal events. Therefore this channel seems promising.

\begin{table}
  \renewcommand{\arraystretch}{1.3}
  \begin{bigcenter}
    \footnotesize
    \begin{tabular}{lccccccc}
      & $HH$ & $b\bar{b}\gamma\gamma$ & $t\bar{t}\gamma\gamma$ &$ZH$ &
       $S/B$ & $S/\sqrt{B}$ \\ \hline
      Cross section NLO [fb] & $8.92\times 10^{-2}$  & $5.05\times
      10^{3}$ & $1.39$ & $3.33\times 10^{-1}$ & $1.77\times 10^{-5}$ &
      $6.87\times 10^{-2}$ \\
      Reconstructed Higgs from $b s$ & $4.37\times 10^{-2}$ &
      $4.01\times 10^{2}$ & $8.70\times 10^{-2}$ & $1.24\times
      10^{-3}$ & $1.09\times 10^{-4}$ & $1.20\times 10^{-1}$ \\
      Reconstructed Higgs from $\gamma s$ & $3.05\times 10^{-2}$ &
      $1.78$ & $2.48\times 10^{-2}$ & $3.73\times 10^{-4}$ &
      $1.69\times 10^{-2}$ & $1.24$ \\
      Cut on \mhh & $2.73\times 10^{-2}$ & $3.74\times 10^{-2}$ &
      $7.45\times 10^{-3}$ & $1.28\times 10^{-4}$ & $6.07\times
      10^{-1}$ & $7.05$ \\
      Cut on $\pth$ & $2.33\times 10^{-2}$ & $3.74\times 10^{-2}$ &
      $5.33\times 10^{-3}$ & $1.18\times 10^{-4}$ & $5.44\times
      10^{-1}$ & $6.17$ \\
      Cut on \etah & $2.04\times 10^{-2}$ & $1.87\times 10^{-2}$ &
      $3.72\times 10^{-3}$ & $9.02\times 10^{-5}$ & $9.06\times
      10^{-1}$ & $7.45$ \\
      Cut on $ \Delta R(b,b)$ & $1.71\times 10^{-2}$ & $0.00$ &
      $3.21\times 10^{-3}$ & $7.44\times 10^{-5}$ & $5.21$ & $16.34$
      \\ \hline
      ``Detector level'' & $1.56\times 10^{-2}$ & $0.00$ & $8.75\times
      10^{-3}$ & $8.74\times 10^{-3}$ & $8.92\times 10^{-1}$ & $6.46$
    \end{tabular}
    \it{\caption{Cross section values of the $HH$ signal and the
        various backgrounds expected at the LHC at $\sqrt{s}=14$~TeV,
        the signal to background ratio $S/B$ and the significance
        $S/\sqrt{B}$ for $\lum =3000$ fb$^{-1}$ in the $\bbgg$ channel
        after applying the cuts discussed in the
        text.\label{table_bbgg}}}
  \end{bigcenter}
\end{table}

A realistic assessment of the prospects for measuring the signal in
the $\bbgg$ final state depends mostly on a realistic simulation of
the diphoton fake rate due to multijet production, which is the
dominant background in such an analysis. Our first parton level study
gives a rough idea of how promising the $\bbgg$ channel is.

In the following we perform a full analysis including fragmentation
and hadronization effects, initial and final state radiations by using
{\tt Pythia 6.4} in order to assess more reliably whether the
promising features of this channel survive in a real experimental
environment. All the events are processed through {\tt
  Delphes}~\cite{Ovyn:2009tx}, the tool which is used for the detector
simulation. For the jet reconstruction we use the anti-$k_T$ algorithm
with a radius parameter of $R=0.5$. We still consider a $b$--tagging
efficiency of $70\%$. We keep the same acceptance cuts as before
except for the transverse momentum of the reconstructed $b$ jet and
photon which we increase up to $p_{T,b/\gamma}>50~{\rm GeV}$. We also
enlarge the window for the reconstructed Higgs boson coming from the
$b$ quark pair, by requiring $100~{\rm GeV} \, < \, M_{b\bar{b}} \,<
\, 135~{\rm GeV}$. We select events with exactly two reconstructed $b$
jets and two photons.

The final result is displayed in the last line of
Table~\ref{table_bbgg}. The final significance $S/\sqrt{B}$ for this
simulation has decreased to 6.5 for an integrated luminosity of
3000 fb$^{-1}$, corresponding to 47 events. Though low, the
significance nevertheless is promising enough to trigger a real
experimental analysis as can be performed only by the experimental
collaborations and which is beyond the scope of this work.

\subsection{The $\bbtt$ decay channel}

The $\bbtt$ decay channel is promising for low mass Higgs boson pair
production at the LHC and has been previously studied in
Ref.~\cite{early, bbtautau}. An important part of this analysis will
depend on the ability to reconstruct the $b$ quark pair and the $\tau$
pair. As the real experimental assessment of such a reconstruction is
beyond the scope of our work we will perform in the following a parton
level analysis, assuming a perfect $\tau$
reconstruction\footnote{There have been improvements over the last
  years to reconstruct the invariant mass of a $\tau$ pair. In
  particular, the use of the Missing Mass Calculator algorithm offers
  very promising results~\cite{mmc}. It is used by experimental
  collaborations at the LHC in the $H\to \tau\bar{\tau}$ search
  channel.}. The analysis thus represents an optimistic estimate of
what can be done at best to extract the trilinear Higgs self--coupling
in the $\bbtt$ channel.

We consider the two QCD--QED continuum background final states 
$\bbtt$ and $b\bar{b}\bar{\tau} \nu_\tau\tau \bar\nu_\tau$ calculated at
tree--level. The $b\bar{b}\bar{\tau} \nu_\tau\tau\bar\nu_\tau$ final state
background dominantly stems from $t\bar{t}$ production with the
subsequent top quark decays $t \rightarrow b W$ and $W \rightarrow
\bar\tau\nu_\tau$. We also include backgrounds coming from single Higgs
production in association with a $Z$ boson and the subsequent decays
$H\rightarrow \tau\bar{\tau}$ and $Z\rightarrow b\bar{b}$ or
$H\rightarrow b\bar{b}$ and $Z\rightarrow \tau\bar{\tau} $. The
effects of QCD corrections are included in our calculation via
multiplicative $K$--factors which are summarized in
Table~\ref{K_bbtt}. All $K$--factors are taken at NLO except for the
single Higgs--strahlung process which is taken at NNLO QCD. The
factorization and renormalization scales have been taken at $M_{HH}$
for the signal and at specific values for each background process.

\begin{table}
  \begin{center}
    \small
    \begin{tabular}{c|cccc}
      $\sqrt{s}$~[TeV] & $HH$ & $b\bar{b}\tau \bar{\tau}$ &
      $t\bar{t}$ & $ZH$ \\ \hline
      14 &  1.88 & 1.21 & 1.35 & 1.33 
    \end{tabular}
    \it{\vspace{-2mm}\caption{$K$--factors for $gg\to HH$,
        $b\bar{b}\tau\bar{\tau}$~\protect\cite{ellis},
        $t\bar{t}$~\protect\cite{lhctop} and $ZH$
        production~\protect\cite{LHCXS} at $\sqrt{s}=14$~TeV. The
        Higgs boson mass is assumed to be
        $M_H=125$~GeV.\label{K_bbtt}}}
  \end{center}
\end{table}

\begin{figure}[!h]
  \begin{bigcenter}
    \begin{tabular}{cc}
      \includegraphics[scale=0.4]{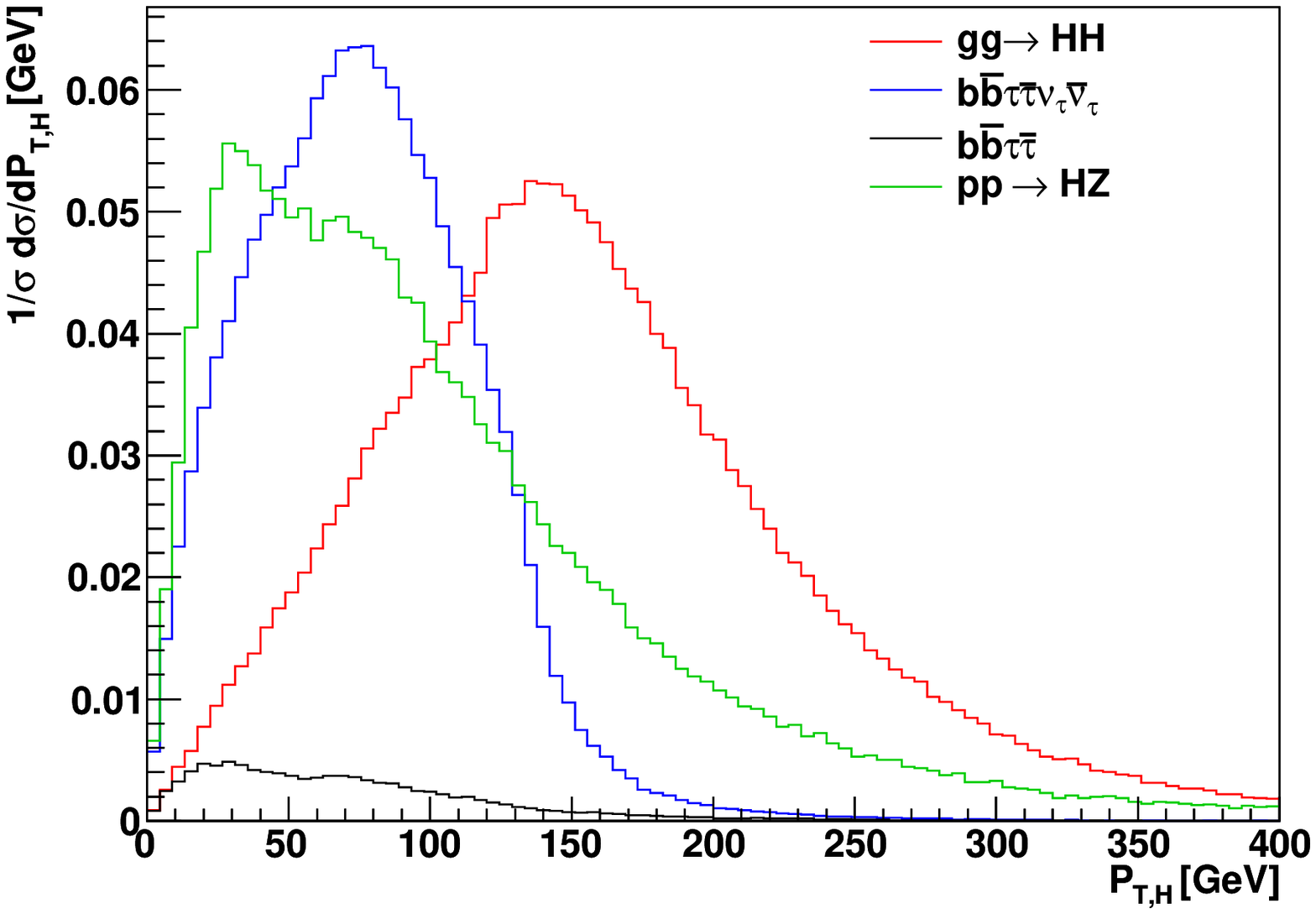}
      &
      \includegraphics[scale=0.4]{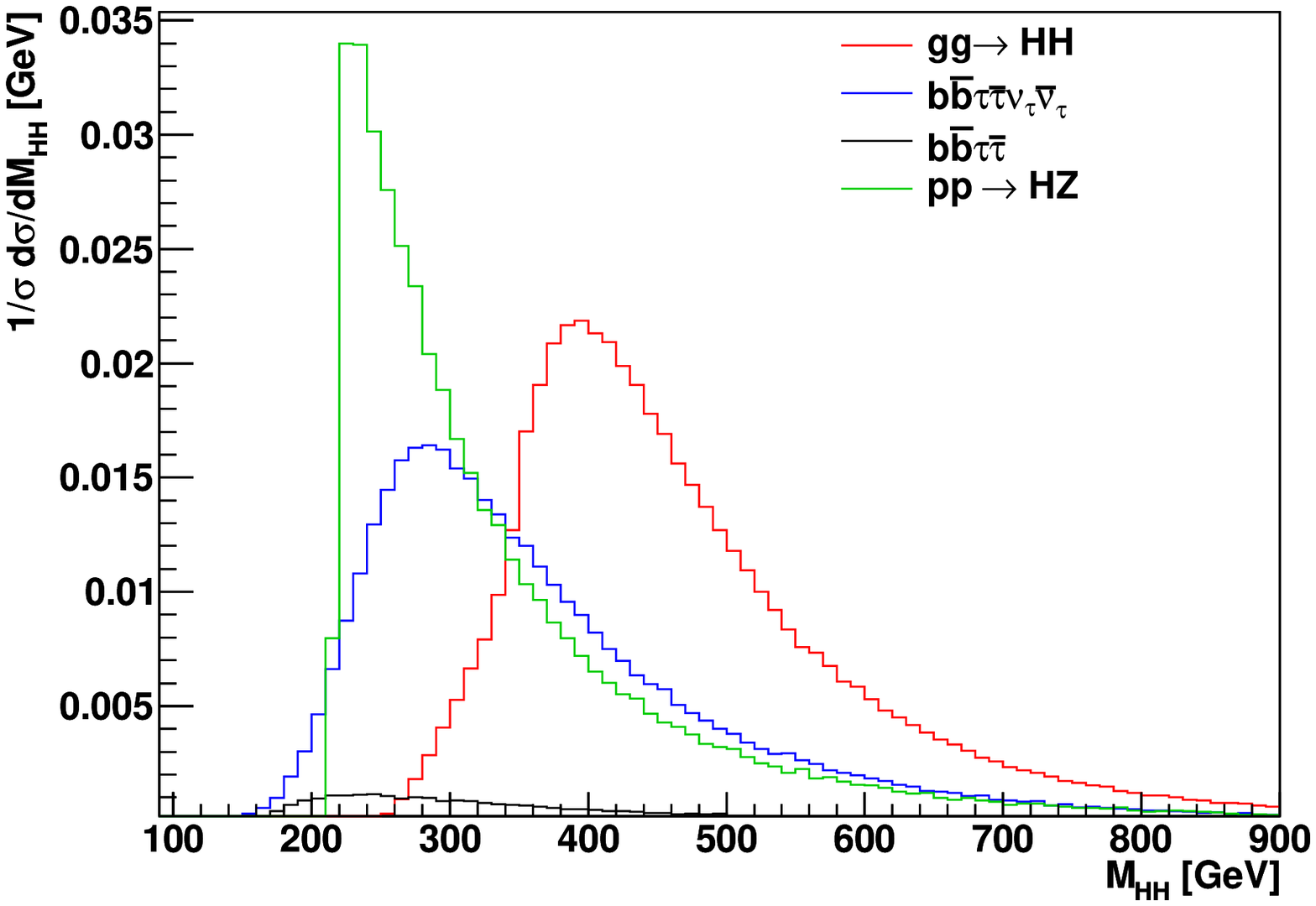}\\
    \end{tabular}
    \includegraphics[scale=0.4]{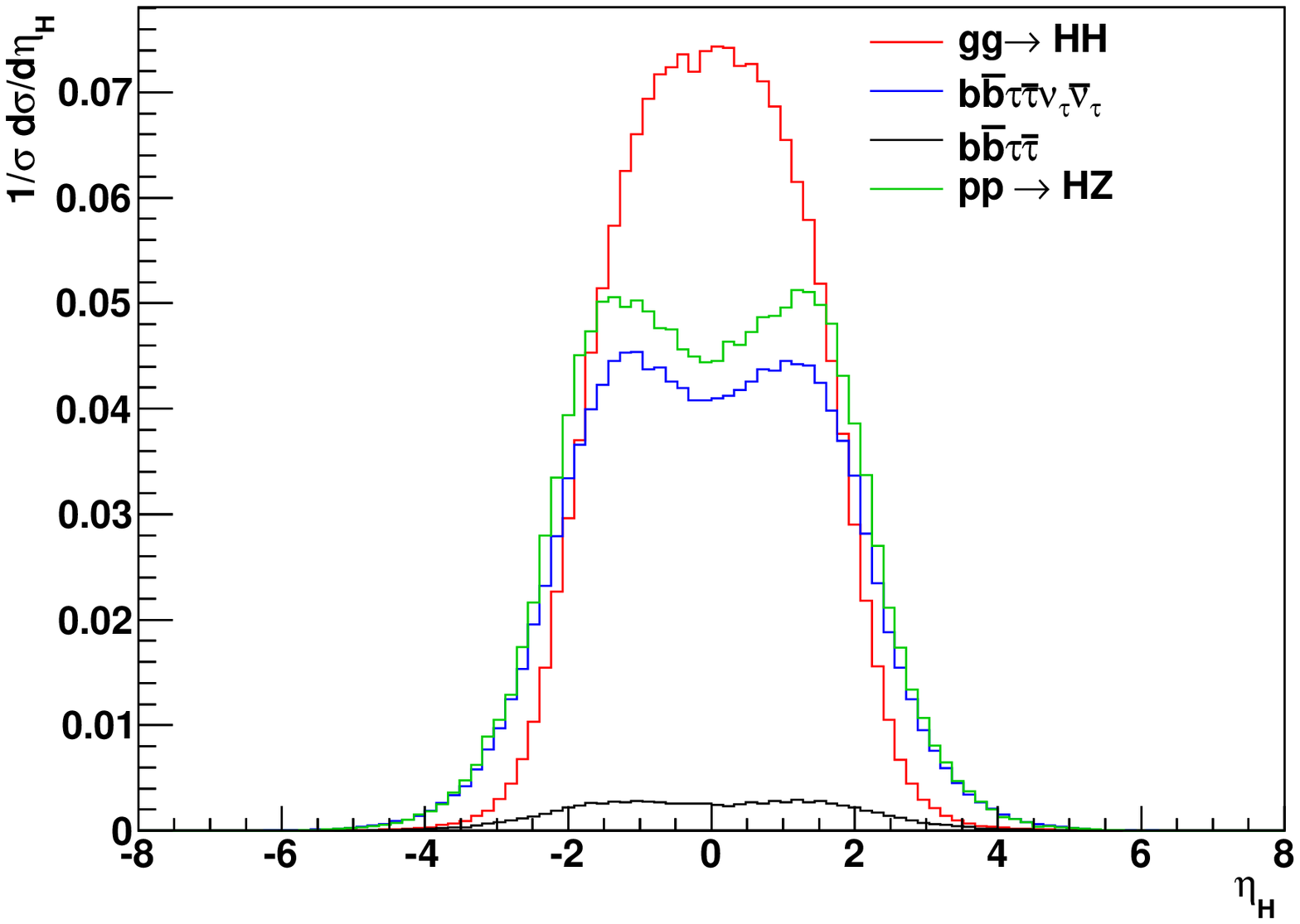}
  \end{bigcenter}
  \it{\vspace{-6mm}\caption{Normalized distributions of $\pth$, $\mhh$
      and $\etah$ for signal and backgrounds in the $\bbtt$
      channel. \label{distributions_bbtt}}}
\end{figure}

Concerning the choice of our cuts, we demand exactly one $b$ quark
pair and one $\tau$ pair. The $b$ quark pair is required to fulfill
$p_{T,b}>30~{\rm GeV}$ and $\vert\eta_b\vert<2.4$. We assume the
$b$--tagging efficiency to be $70\%$ and the $\tau$--tagging
efficiency to be $50\%$ and we neglect fake rates in both cases. The
$\tau$ pair has to fulfill $p_{T,\tau}>30~{\rm GeV}$ and
$\vert\eta_\tau\vert<2.4$. The reconstructed Higgs boson from the $b$
quark pair is required to reproduce the Higgs mass within a window of
25~GeV, $112.5~{\rm GeV} \, < \, M_{b\bar{b}} \,< \, 137.5~{\rm
  GeV}$. The reconstructed Higgs boson from the $\tau$ pair needs to
reproduce the Higgs mass within a window of 50~GeV, $100~{\rm GeV} \,
< \, M_{\tau\bar{\tau}} \,< \, 150~{\rm GeV}$ or within a window of
25~GeV, $112.5~{\rm GeV} \, < \, M_{\tau\bar{\tau}} \,< \, 137.5~{\rm
  GeV}$, in a more optimistic scenario. In addition to these
acceptance cuts we also add more advanced cuts for this parton level
analysis, based on the distributions shown in
Fig.~\ref{distributions_bbtt} and in a similar way as what has been
done in the previous $b\bar{b} \gamma\gamma$ analysis. We first demand
the invariant mass of the reconstructed Higgs pair to fulfill $\mhh >$
350 GeV. In addition, we remove events which do not satisfy $\pth >$
100 GeV. We do not use a cut on the pseudorapidity of the
reconstructed Higgs bosons in this analysis as it would reduce the
significance.
 
The different results of our parton level analysis are summarized in
Table~\ref{table_bbtt}. The cuts allow for an improvement of two
orders of magnitude in the significance $S/\sqrt{B}$. The final
$S/\sqrt{B}$ is $6.71$ for an integrated luminosity of $\lum=3000$
fb$^{-1}$, corresponding to 330 signal events. We then conclude that
this channel is promising. In the last line we reproduce our result
for the optimistic requirement of $112.5~{\rm GeV} \, < \,
M_{\tau\tau} \, < \, 137.5~{\rm GeV}$ leading to the final
significance $S/\sqrt{B}=9.36$ for an integrated luminosity of 3000
fb$^{-1}$. Already for a planned mid--term integrated luminosity of
300 fb$^{-1}$ at 14 TeV the expectations are promising with 33 signal
events and a significance $S/\sqrt{B} = 2.96$ in the optimistic
scenario.

\begin{table}
  \renewcommand{\arraystretch}{1.3}
  \begin{bigcenter}
    \footnotesize
    \begin{tabular}{lccccccc}
      & $HH$ & $b\bar{b}\tau\bar{\tau}$ &
      $b\bar{b}\tau\bar{\tau}\nu_{\tau}\bar{\nu}_{\tau}$ & $ZH$ &
      $S/B$ & $S/\sqrt{B}$ \\ \hline
      Cross section NLO [fb] & $2.47$ & $2.99\times 10^{4}$ &
      $8.17\times 10^{3}$ & $2.46\times 10^{1}$ & $6.48\times 10^{-5}$
      & $6.93\times 10^{-1}$ \\
      Reconstructed Higgs from $\tau s$ & $2.09\times 10^{-1} $ &
      $8.35\times 10^{1}$ & $1.58\times 10^{2}$ & $5.70\times 10^{-1}$
      & $8.63\times 10^{-4}$ & $7.36\times 10^{-1}$\\
      Reconstructed Higgs from $b s$ & $1.46\times 10^{-1} $ &
      $6.34\times 10^{-1}$ & $1.43\times 10^{1}$ & $3.75\times
      10^{-2}$ & $9.75\times 10^{-3}$ & $2.07$ \\
      Cut on \mhh & $1.30\times 10^{-1}$ & $1.37\times 10^{-1}$ &
      $1.74$ & $1.26\times 10^{-2}$ & $6.88\times 10^{-2}$ &
       $5.18$ \\
      Cut on \pth & $1.10\times 10^{-1}$ & $7.80\times 10^{-2}$ &
      $7.17\times 10^{-1}$ & $1.15\times 10^{-2}$ & $1.36\times
      10^{-1}$ & $6.71$ \\ \hline
      With $112.5~{\rm GeV}\, <\,M_{\tau\bar{\tau}} \,<\,137.5~{\rm
        GeV}$ & $1.10\times 10^{-1}$ & $3.41\times 10^{-2}$ &
      $3.76\times 10^{-1}$ & $3.15\times 10^{-3}$ & $2.67\times
      10^{-1}$ & $9.37$
    \end{tabular}
    \it{\vspace{-1mm}\caption{Cross section values of the of $HH$
        signal and the various backgrounds expected at the LHC at
        $\sqrt{s}=14$~TeV, the signal to background ratio $S/B$ and the
        significance $S/\sqrt{B}$ for $\lum =3000$ fb$^{-1}$ in the
        $\bbtt$ channel after applying the cuts discussed in the
        text.\label{table_bbtt}}}
  \end{bigcenter}
\end{table}

\subsection{The $\bbww$ decay channel}

The analysis in this channel is difficult as the leptonic $W$ boson
decays lead to missing energy in the final state. Consequently, one of
the two Higgs bosons cannot {\it a priori} be reconstructed equally
well as the other Higgs boson, thus reducing our capability to
efficiently remove the background with the canonical acceptance cuts
previously applied in the other decay channels. This channel with one
lepton plus jets final state has been studied
in~\cite{bbtautau,bbWW}. We only consider here the decay $W\rightarrow
\ell \nu_{\ell}$ ($\ell=e,\mu$) with a branching ratio of $10.8\%$.

We take into account the continuum background which contains all
processes with the $b\bar{b} \ell \nu_{\ell} \ell \nu_{\ell}$ final
states at tree--level, for example $qq/gg \to b^* \bar{b}^* \to \gamma
b Z \bar b \to b\bar{b} \ell \ell Z$ with the subsequent splitting
$Z\to \nu_{\ell} \bar{\nu_{\ell}}$. We proceed in a similar manner as
in the previous analyses. We generate the signal and the backgrounds
with the following parton-level cuts. We require that the $b$ quarks
fulfill $p_{T,b}>30~{\rm GeV}$ and $\vert\eta_b\vert<2.4$. We consider
the $b$--tagging efficiency to be $70\%$. The leptons have to fulfill
$p_{T,\ell}>20~{\rm GeV}$ and $\vert\eta_\ell\vert<2.4$. The
reconstructed Higgs boson from the $b$ quark pair has to reproduce the
Higgs boson mass within a window of 25~GeV, $112.5~{\rm GeV} \, < \,
M_{b\bar{b}} \,< \, 137.5~{\rm GeV}$. We also require that the missing
transverse energy respects $E_{T}^{miss} > 20~{\rm GeV}$.

\begin{figure}[!h]
  \begin{bigcenter}
    \begin{tabular}{cc}
      \includegraphics[scale=0.4]{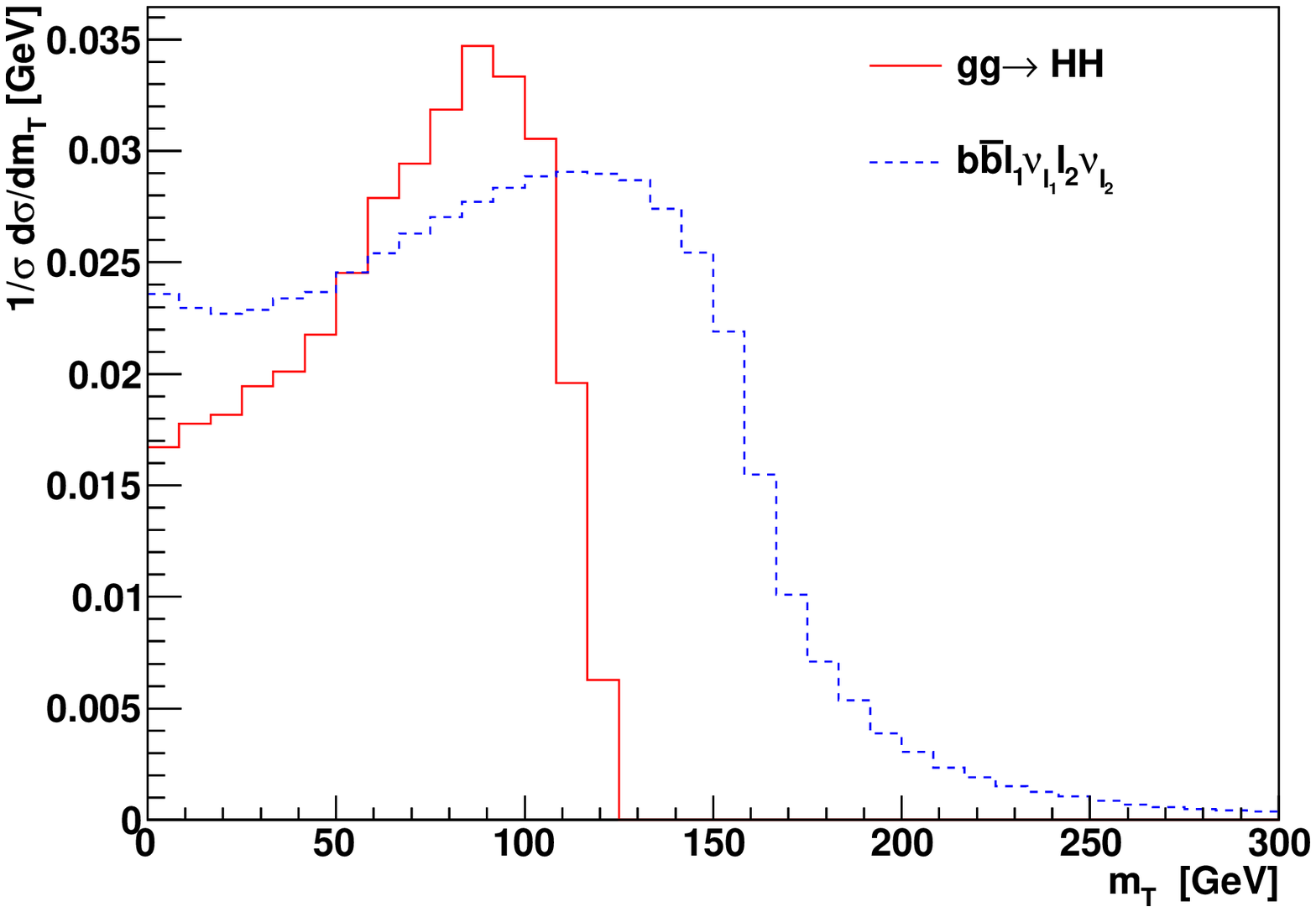} 
      &
      \includegraphics[scale=0.4]{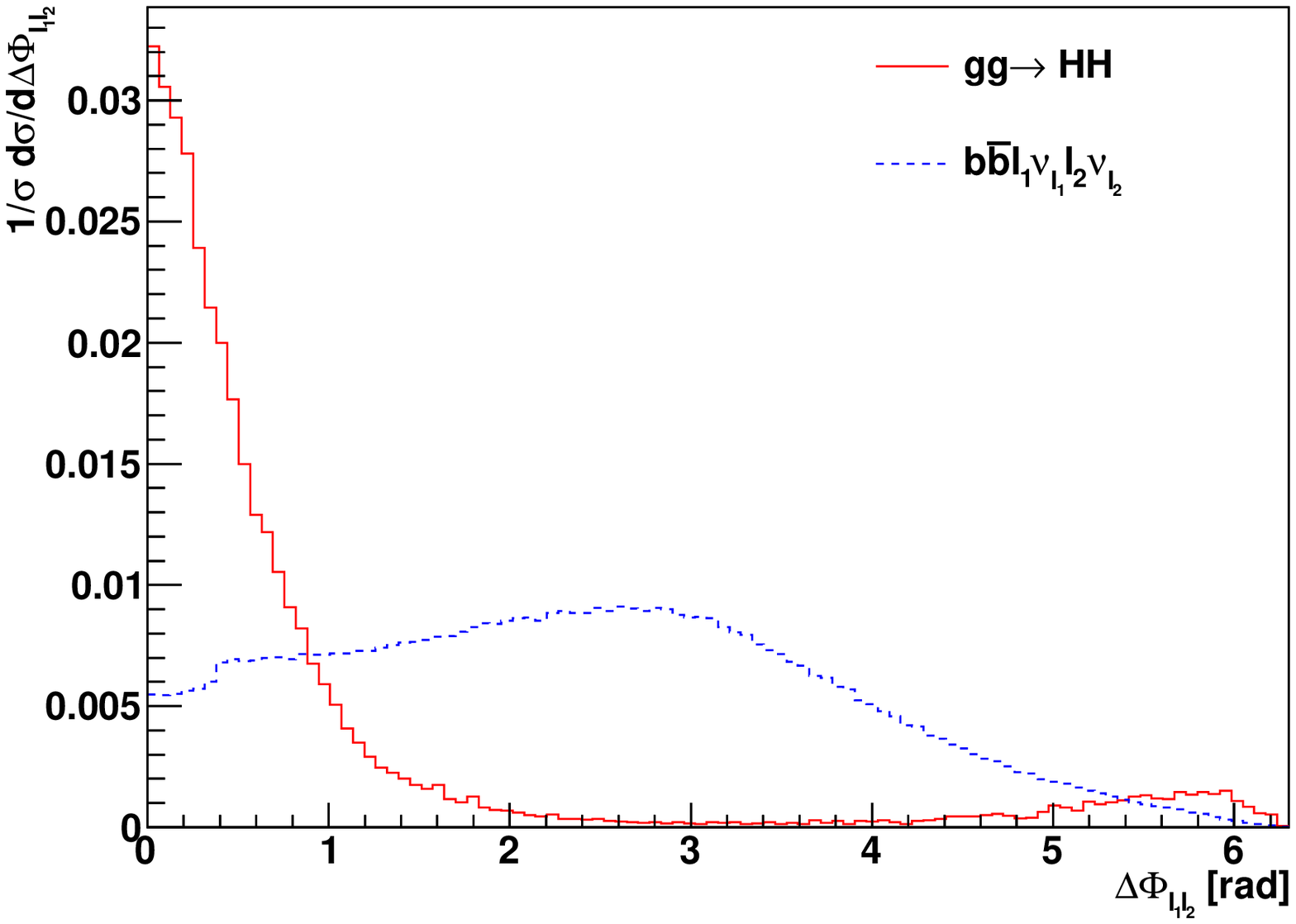}\\
    \end{tabular}
    \includegraphics[scale=0.4]{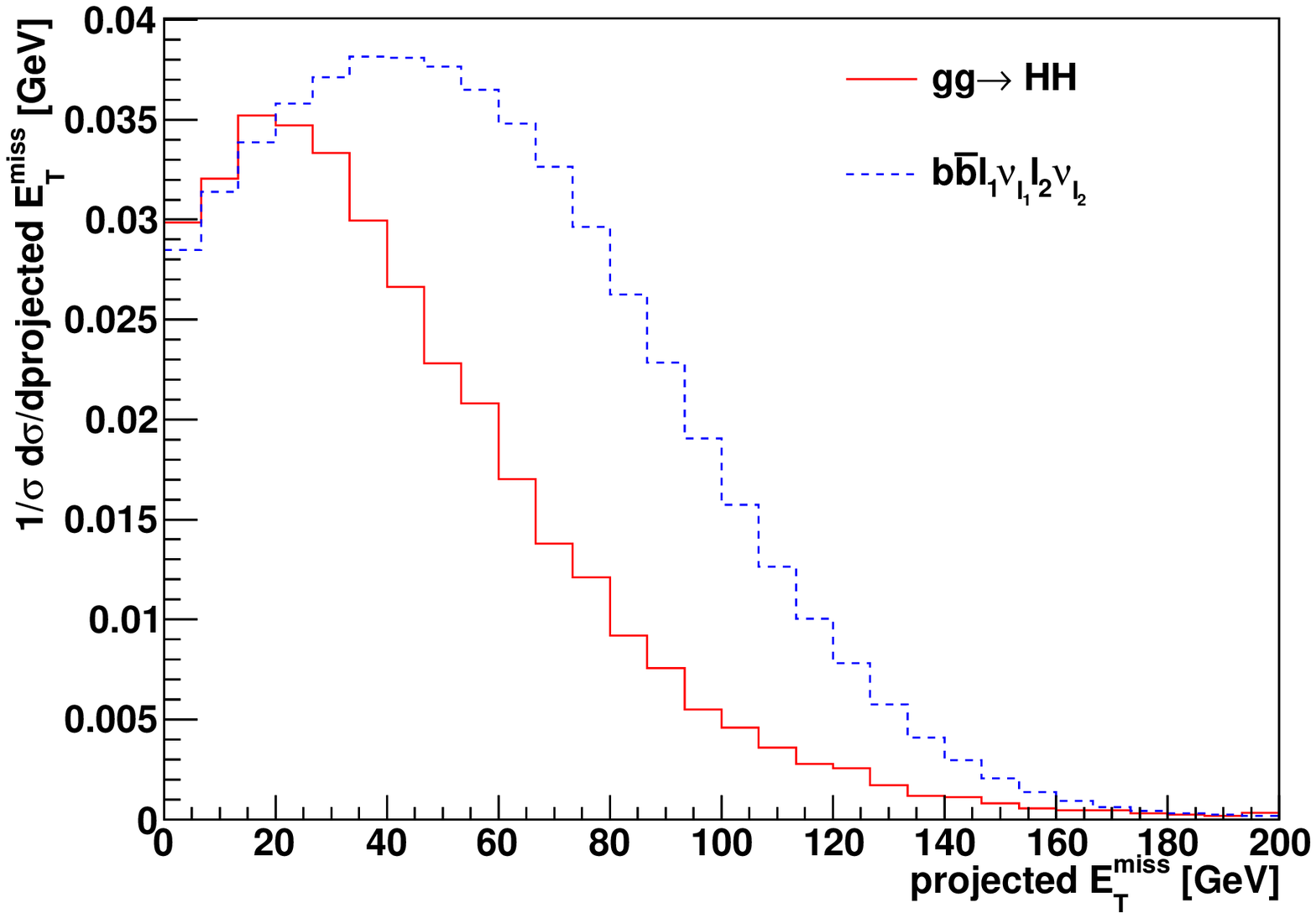}
  \end{bigcenter}
  \it{\vspace{-6mm}\caption{Normalized distributions of $M_{T}$, $\Delta
      \phi_{l_{1}l_{2}}$ and projected missing transverse energy
      $\tilde{E}_{T}^{miss}$ for signal and background channels in the
      $b\bar{b}  l_{1}\nu_{l_{1}} l_{2}\nu_{l_{2}}$ final states of
      the $\bbww$ channel.\label{distributions_bbww}}}
\end{figure}

As done in the previous subsections, we also add more advanced cuts
for this parton level analysis, based on the distributions shown in
Fig.~\ref{distributions_bbww}. The distributions on the upper left of
Fig.~\ref{distributions_bbww} correspond to the transverse mass of the
lepton pair, being defined as $M_{T}=\sqrt{2 p_{T}^{\ell \ell}
  E_{T}^{miss} (1-\cos \Delta \phi(E_{T}^{miss},\ell\ell ))}$, where
$\Delta \phi(E_{T}^{miss},\ell\ell )$ is the angle between the
missing transverse momentum and the transverse momentum of the
dilepton system. The distribution of the signal has an endpoint at the
value of $\mh$. The distributions on the upper right of
Fig.\ref{distributions_bbww} represent the angle between the two
leptons projected on the transverse plane,
$\Delta\phi_{\ell_{1}\ell_{2}}$. The angle is reduced for the signal
compared to the broad distribution of the background. The last
distributions on the bottom of Fig.~\ref{distributions_bbww} display
the projected missing transverse energy $\tilde{E}_{T}^{miss} =
E_{T}^{miss} \sin \Delta \phi(E_{T}^{miss}, \Lep)$ for $\Delta
\phi(E_{T}^{miss}, \Lep) \leq \pi/2$, where $\Delta \phi(E_{T}^{miss},
\Lep)$ is the angle between the missing transverse momentum and the
transverse momentum of the nearest lepton candidate. If $\Delta
\phi(E_{T}^{miss}, \Lep) > \pi/2$, then $\tilde{E}_{T}^{miss} =
E_{T}^{miss}$. The signal distribution is shifted to the left compared
to the background distribution.

We first require the transverse mass of the lepton pair to be $M_{T} <
125~{\rm GeV}$. We then remove events which do not satisfy $\Delta
\phi_{\ell_{1}\ell_{2}} < 1.2$ and we also add a constraint on the angle
between the two leptons, $\Delta \theta_{\ell_{1}\ell_{2}} < 1.0$. We
demand the missing transverse energy to fulfill $E_{T}^{miss} >
120~\rm GeV$ and the projected energy to satisfy $\tilde{E}_{T}^{miss}
< 80~\rm GeV$. Note that the $\tilde{E}_{T}^{miss}$ distribution
displayed in Fig.~\ref{distributions_bbww} is obtained after the
acceptance cuts having been applied but before the advanced cuts. The
cuts on $M_T$, $\Delta\phi_{\ell_{1}\ell_{2}}$,
$\Delta\theta_{\ell_{1}\ell_{2}}$ and $\tilde{E}_T^{miss}$ modify this
distribution and explain why the $\tilde{E}_T^{miss}$ cut, which would
seem not to be efficient, actually improves the significance.

\begin{table}[!h]
  \renewcommand{\arraystretch}{1.3}
  \begin{bigcenter}
    \small
    \begin{tabular}{lcccccc}
      & $HH$ & $b\bar{b}  l_{1}\nu_{l_{1}} l_{2}\nu_{l_{2}}$ &
      $S/B$ & $S/\sqrt{B}$ \\ \hline
      Cross section NLO [fb] & $3.92\times 10^{-1}$ & $2.41\times
      10^{4}$ & $1.63\times 10^{-5}$ & $1.38 \times 10^{-1}$ \\
      Reconstructed Higgs from $b s$ & $6.18\times 10^{-2}$ &
      $1.89\times 10^{2}$ & $3.27\times 10^{-4}$ & $2.46\times
      10^{-1}$ \\
      Cut on $M_{T}$ & $6.18.\times 10^{-2}$ & $1.19\times 10^{2}$ &
      $5.19\times 10^{-4}$ & $3.10\times 10^{-1}$ \\
      Cut on $\Delta \phi_{\ell_{1}\ell_{2}}$ & $5.37\times 10^{-2}$ &
      $6.96\times 10^{1}$ & $7.72\times 10^{-4}$ & $3.53\times
      10^{-1}$ \\
      Cut on $\Delta \theta_{\ell_{1}\ell_{2}}$ & $5.17\times 10^{-2}$ &
      $5.65\times 10^{1}$ & $9.15\times 10^{-4}$ & $3.77\times
      10^{-1}$ \\
      Cut on $E_{T}^{miss}$ & $8.41\times 10^{-3}$ & $3.77\times
      10^{-1}$ & $2.22\times 10^{-2}$ & $7.50\times 10^{-1}$ \\
      Cut on $\tilde{E}_{T}^{miss}$ & $4.59\times 10^{-3}$ &
      $2.70\times 10^{-2}$ & $1.70\times 10^{-1}$ & $1.53$
    \end{tabular}
    \it{\vspace{-1mm}\caption{Cross section values of the $HH$ signal
        and the considered background expected at the LHC at
        $\sqrt{s}=14$~TeV, the signal to background ratio $S/B$ and the
        significance $S/\sqrt{B}$ for \lum =3000 fb$^{-1}$ in the
        $\bbww$ channel after applying the cuts discussed in the
        text.}\label{table_bbww}}
  \end{bigcenter}
\end{table}

The results for the LHC at $\sqrt{s}=14$~TeV are summarized in
Table~\ref{table_bbww}. While the cuts allow for an improvement of the
significance $S/\sqrt{B}$ by about one order of magnitude, we are
still left with a very small signal to background ratio. Thus, this
channel using the final states considered here is not very promising.


\section{Conclusions} 

In this paper we have discussed in detail the main Higgs pair
production processes at the LHC, gluon fusion, vector boson fusion,
double Higgs--strahlung and associated production with a top quark
pair. They allow for the determination of the trilinear Higgs
self--coupling $\lambda_{HHH}$, which represents a first important step
towards the reconstruction of the Higgs potential and thus the final
verification of the Higgs mechanism as the origin of electroweak
symmetry breaking. We have included the important QCD corrections at
NLO to gluon fusion and vector boson fusion and calculated for the
first time the NNLO corrections to double Higgs--strahlung. It turns
out that the gluon initiated process to $ZHH$ production  which
contributes at NNLO is sizeable in contrast to the single
Higgs-strahlung case. We have discussed in detail the various
uncertainties of the different processes and provided numbers for the
cross sections and the total uncertainties at four c.m. energies, i.e. 8,
14, 33 and 100 TeV. It turns out that they are of the
  order of $40\%$ in the gluon fusion channel while they are much more
limited in the vector bosons fusion and double Higgs--strahlung
processes, i.e. below $10\%$. Within the SM we also studied the
sensitivities of the double Higgs production processes to the
trilinear Higgs self--coupling in order to get an estimate of how
accurately the cross sections have to be measured in order to extract
the Higgs self--interaction with sufficient accuracy.

In the second part of our work we have performed a parton level
analysis for the dominant Higgs pair production process through gluon
fusion in different final states which are $b\bar{b}\gamma\gamma$,
$b\bar{b}\tau\bar{\tau}$ and $b\bar{b}W^+W^-$ with the $W$ bosons decaying
leptonically. Due to the smallness of the signal and the large QCD
backgrounds the analysis is challenging. The $b\bar{b}W^+W^-$ final
state leads to a very small signal to background ratio after applying
acceptance and selection cuts so that it is not promising. On the
other hand, the significances obtained in the $b\bar{b}\gamma\gamma$
and $b\bar{b}\tau\bar{\tau}$ final states after cuts are $\sim 16$ and
$\sim 9$, respectively, with not too small event numbers. They are
thus promising enough to start a real experimental analysis taking
into account detector and hadronization effects, which is beyond the
scope of our work. Performing a first simulation on the detector level
for the $b\bar{b}\gamma\gamma$ state shows, however, that the
prospects are good in case of high luminosities. Taking into account
theoretical and statistical uncertainties and using the sensitivity
plot, Fig.~\ref{sensitivity}, the trilinear Higgs self-coupling
$\lambda_{HHH}$ can be expected to be measured within a factor of
two.

\newpage

\noindent
{\bf Acknowledgements}:

\noindent
We are grateful to G. Salam, C. Grojean and M. Spannowsky for
discussions as well as S. Moretti for clarifications.  J.B. and
R.G. also thank S. Palmer, M. Rauch and D. N. Le for clarifications
and discussions about the VBF and Higgs--strahlung processes. A.D. and
J.Q. thank CERN for its kind hospitality. J.B., R.G. and
M.M.M. acknowledge the support from the Deutsche
Forschungsgemeinschaft via the Sonderforschungsbereich/Transregio
SFB/TR-9 Computational Particle Physics. R.G. acknowledges the
financial support from the Landesgraduiertenkolleg. A.D. is supported
by the ERC advanced grant Higgs@LHC.

\end{document}